\documentclass[useAMS,usenatbib]{mn2e}
\usepackage{natbib}
\usepackage{amssymb,amsmath,amsfonts}
\usepackage{epsfig}
\usepackage{mathptmx}
\usepackage{longtable}
\usepackage{graphicx}
\usepackage{comment}
\usepackage[caption=false]{subfig}
\usepackage{color}

\def\reff@jnl#1{{\rm#1\/}}

\def\aj{\reff@jnl{AJ}}                  
\def\araa{\reff@jnl{ARA\&A}}            
\def\apj{\reff@jnl{ApJ}}                        
\def\apjl{\reff@jnl{ApJ}}               
\def\apjs{\reff@jnl{ApJS}}              
\def\ao{\reff@jnl{Appl.Optics}}         
\def\apss{\reff@jnl{Ap\&SS}}            
\def\aap{\reff@jnl{A\&A}}               
\def\aapr{\reff@jnl{A\&A~Rev.}}         
\def\aaps{\reff@jnl{A\&AS}}             
\def\azh{\reff@jnl{AZh}}                        
\def\baas{\reff@jnl{BAAS}}              
\def\jrasc{\reff@jnl{JRASC}}            
\def\memras{\reff@jnl{MmRAS}}           
\def\mnras{\reff@jnl{MNRAS}}            
\def\pra{\reff@jnl{Phys. Rev. A}}         
\def\prb{\reff@jnl{Phys. Rev. B}}         
\def\prc{\reff@jnl{Phys. Rev. C}}         
\def\prd{\reff@jnl{Phys. Rev. D}}         
\def\prl{\reff@jnl{Phys. Rev. Lett}}      
\def\pasp{\reff@jnl{PASP}}              
\def\pasj{\reff@jnl{PASJ}}              
\def\qjras{\reff@jnl{QJRAS}}            
\def\skytel{\reff@jnl{S\&T}}            
\def\solphys{\reff@jnl{Solar~Phys.}}    
\def\sovast{\reff@jnl{Soviet~Ast.}}     
\def\ssr{\reff@jnl{Space~Sci.Rev.}}     
\def\zap{\reff@jnl{ZAp}}                        
\def\nat{\reff@jnl{Nature}}             

\def\p#1by#2{{\partial{#1} \over \partial{#2}}}
\def\pp#1by#2#3{{\partial^2{#1} \over \partial{#2}\partial{#3}}}
\def\d#1by#2{{{\rm d}{#1} \over {\rm d}{#2}}}
\def\dd#1by#2#3{{{\rm d}^2{#1} \over {\rm d}{#2}{\rm d}{#3}}}

\title[]
{A joint analysis of AMI and CARMA observations of the recently discovered SZ galaxy cluster system AMI-CL J0300+2613}

\label{firstpage}

\author[Shimwell et~al.]{AMI Consortium:
 Timothy W. Shimwell$^1$\thanks{E-mail: Timothy.Shimwell@csiro.au},
  John M. Carpenter$^{2}$, 
  Farhan Feroz$^{3}$,
   \newauthor
  Keith J. B. Grainge$^{3,4}$,
  Michael P. Hobson$^{3}$, 
  Natasha Hurley-Walker$^5$,
   \newauthor
  Anthony N. Lasenby$^{3,4}$,
  Malak Olamaie$^3$,
 Yvette C. Perrott$^3$,
 Guy G. Pooley$^3$,
 \newauthor
 Carmen Rodr{\'i}guez-Gonz{\'a}lvez$^6$,
 Clare Rumsey$^3$,
 Richard D. E. Saunders$^{3,4}$,
  \newauthor
 Michel P. Schammel$^3$,
 Paul F. Scott$^3$,
 David J. Titterington$^3$,
 Elizabeth M. Waldram$^3$\\
 $^1$ CSIRO Astronomy \& Space Science, Australia Telescope National Facility, PO Box 76, Epping, NSW 1710, Australia \\
 $^2$ Department of Astronomy, California Institute of Technology, 1200 E. California Blvd., MC 249-17, Pasadena, CA 91125, USA \\
 $^3$ Astrophysics Group, Cavendish Laboratory, J J Thomson Avenue, Cambridge CB3 0HE\\
 $^4$ Kavli Institute for Cosmology Cambridge, Madingley Road, Cambridge, CB3 0HA\\
 $^5$ International Centre for Radio Astronomy Research, Curtin Institute of Radio Astronomy, 1 Turner Avenue, Technology Park, Bentley, WA 6845, Australia \\
 $^6$ Spitzer Science Center, MS 220-6, California Institute of Technology, Pasadena, CA 91125, USA \\
}

\begin{document}

\maketitle

\begin{abstract}
\noindent


We present Combined Array for Research in Millimeter-wave Astronomy (CARMA) observations of a massive galaxy cluster discovered in the Arcminute Microkelvin Imager (AMI) blind Sunyaev-Zel'dovich (SZ) survey. 
Without knowledge of the cluster redshift a Bayesian analysis of the AMI, CARMA and joint AMI \& CARMA $uv$-data is used to quantify the detection significance and parameterise both the physical and observational properties of the cluster whilst accounting for the statistics of primary CMB anisotropies, receiver noise and radio sources. The joint analysis of the AMI \& CARMA $uv$-data was performed with two parametric physical cluster models: the $\beta$-model; and the model described in \citealt{Olamaie_2012} with the pressure profile fixed according to \cite{Arnaud_2010}. The cluster mass derived from these different models is comparable but our Bayesian evidences indicate a preference for the $\beta$-profile which we, therefore, use throughout our analysis. From the CARMA data alone we obtain a formal Bayesian probability of detection ratio of 12.8:1 when assuming that a cluster exists within our search area; alternatively assuming that \citealt{Jenkins_2001} accurately predicts the number of clusters as a function of mass and redshift, the formal Bayesian probability of detection is 0.29:1. From the Bayesian analysis of the AMI or AMI \& CARMA data the probability of detection ratio exceeds 4.5$\times10^{3}$:1. Performing a joint analysis of the AMI \& CARMA data with a physical cluster model we derive the total mass internal to $r_{200}$ as $M_{T,200}= 4.1\pm 1.1 \times10^{14}M_{\odot}$. Using a phenomenological $\beta$-model to quantify the temperature decrement as a function of angular distance we find a central SZ temperature decrement of 170$\pm24 \mu$K in the AMI \& CARMA data. The SZ decrement in the CARMA data is weaker than expected and we speculate that this is a consequence of the cluster morphology. In a forthcoming study the pipeline that we have developed for the analyses of these data will be used to thoroughly assess the impact of cluster morphology on the SZ decrements that are observed with interferometers such as AMI and CARMA.


\end{abstract}


\section{Introduction}

The Sunyaev-Zel'dovich (SZ) effect occurs when cosmic microwave background (CMB) photons are
inverse-Compton scattered by hot plasma (which dominates the baryon content of the cluster)
in a cluster's gravitational potential well. At frequencies $<217$\,GHz, the scattering causes a decrement in the intensity of the CMB
in the direction of the cluster whilst at higher frequencies an increment is produced (\citealt{Sunyaev_1972}; \citealt{Birkinshaw_1999} and \citealt{Carlstrom_2002}). The SZ surface brightness is proportional to the line-of-sight integral of the plasma
pressure (known as the Comptonization y parameter) and has no dependence on the distance to the cluster. The total SZ flux of a cluster is $\propto d_{A}^{-2} \int n_{e}T_{e} dV$, where $d_{A}$ is the angular diameter distance, $n_e$ is the number density of electrons, $T_{e}$ is the temperature of the electrons and the integral is over the cluster volume. With the assumption that the cluster is virialized, isothermal, spherical, and that all kinetic energy in it is plasma internal energy,  $T_{e} \propto M^{2/3}$, where $M$ is the cluster mass, and the integrated SZ effect is $\propto {M^{5/3}}d_{A}^{-2}$. Hence, the SZ signal is an excellent proxy of mass and blind SZ surveys can provide close to mass-limited samples of galaxy clusters. 
Such samples of clusters are sensitive probes of the formation and evolution of the large-scale structure in the Universe and are used to measure fundamental cosmological parameters such as the rms mass fluctuation in spheres of present radius 8$h^{-1}$Mpc, $\sigma_{8}$, where $h$ is the Hubble parameter.

It is essential to validate low significance detections of galaxy clusters discovered through blind SZ surveys with a different instrument than that used to conduct the survey. SZ blind surveys by the Atacama Cosmology Telescope (ACT), Arcminute Microkelvin Imager (AMI), South Pole Telescope (SPT), Combined Array for Research in Millimeter-wave Astronomy (CARMA) and Planck have produced initial results (e.g. \citealt{Hincks_2010}, \citealt{Menanteau_2010}, \citealt{Staniszewski_2009}, \citealt{Shimwell_2012}, \citealt{Muchovej_2011}, \citealt{Vanderlinde_2009}, \citealt{High_2010} and  \citealt{Planck_2011}) and many of the discovered clusters are now confirmed.  Optical or X-ray observations could be used to validate clusters; however, given that the redshift of blindly detected SZ clusters cannot be determined solely from the SZ observations, the integration time required for such observations is uncertain. Instead, SZ observations with a different instrument can confirm the cluster and the integration time required is much less dependent on redshift. Further SZ observations can be used to compare SZ derived cluster parameters, such as the cluster mass and shape, which is especially important given the recently discovered discrepancies between the AMI and Planck derived cluster parameters (see \citealt{Brown_2012}).

We have used the eight 3.5\,m dishes of CARMA to obtain deep 31\,GHz observations towards the galaxy cluster AMI-CL J0300+2613 
which was discovered in a preliminary analysis of a small region in the AMI blind SZ survey (\citealt{Shimwell_2012}, hereafter Paper I). These are the first targeted follow-up observations towards this cluster with a different telescope. Here we present an analysis of the newly acquired CARMA data and for the first time we conduct a joint analysis of the AMI and CARMA data. For comparison with these new results we include an analysis of the original AMI data used in Paper I.

Hereafter we assume a concordance $\rm{\Lambda}$CDM cosmology, with
$\rm{\Omega_{m}}$ = 0.3, $\rm{\Omega_\Lambda}$ = 0.7 and H$_{0}$ = 70 km\,s$^{-1}$Mpc$^{-1}$.
Coordinates are J2000.

\section{The AMI Blind Cluster Survey}

AMI is a pair of aperture-synthesis interferometric arrays operating at 15.7\,GHz  and is primarily used for imaging the SZ effect. The AMI Large Array (LA) and the AMI Small Array (SA) have identical electronics. The LA, with longer baselines and larger dishes, has higher flux sensitivity and resolution; whereas the SA, with higher surface brightness sensitivity, is ideal for observations of arcminute scale structures such as galaxy clusters.  For SZ observations the LA is used to characterise contaminating sources which can then be subtracted from the SA data. AMI is described in detail
in \cite{Zwart_2008} and the technical aspects of the arrays are
summarised in Table \ref{tab:telescope_tech}.

The AMI data are calibrated and flagged using the in-house software
package \textsc{reduce} and imaging is performed in \textsc{aips}\footnote{http://aips.nrao.edu/}. Further details on the data
reduction and mapping of the AMI blind SZ survey data are provided in Paper I.

The AMI cluster survey has a natural angular resolution of $\sim 3'$  and spans twelve separate deg$^2$ regions that contain no previously recorded massive clusters.  The AMI cluster survey aims to detect SZ-effect signals from clusters of galaxies with a mass above $M_{T,200}$ $=$ 2 $\times$ $10^{14}$$M_{\odot}$ (at z $>$0.2), where
$M_{T,200}$ corresponds to the total cluster mass within a spherical volume such that
the mean interior density is 200 times the mean density of the Universe at the cluster redshift -- the radius of this volume is $r_{200}$.

\begin{figure*}
\centerline{\includegraphics[width= 8.0cm,clip=,angle=0.]{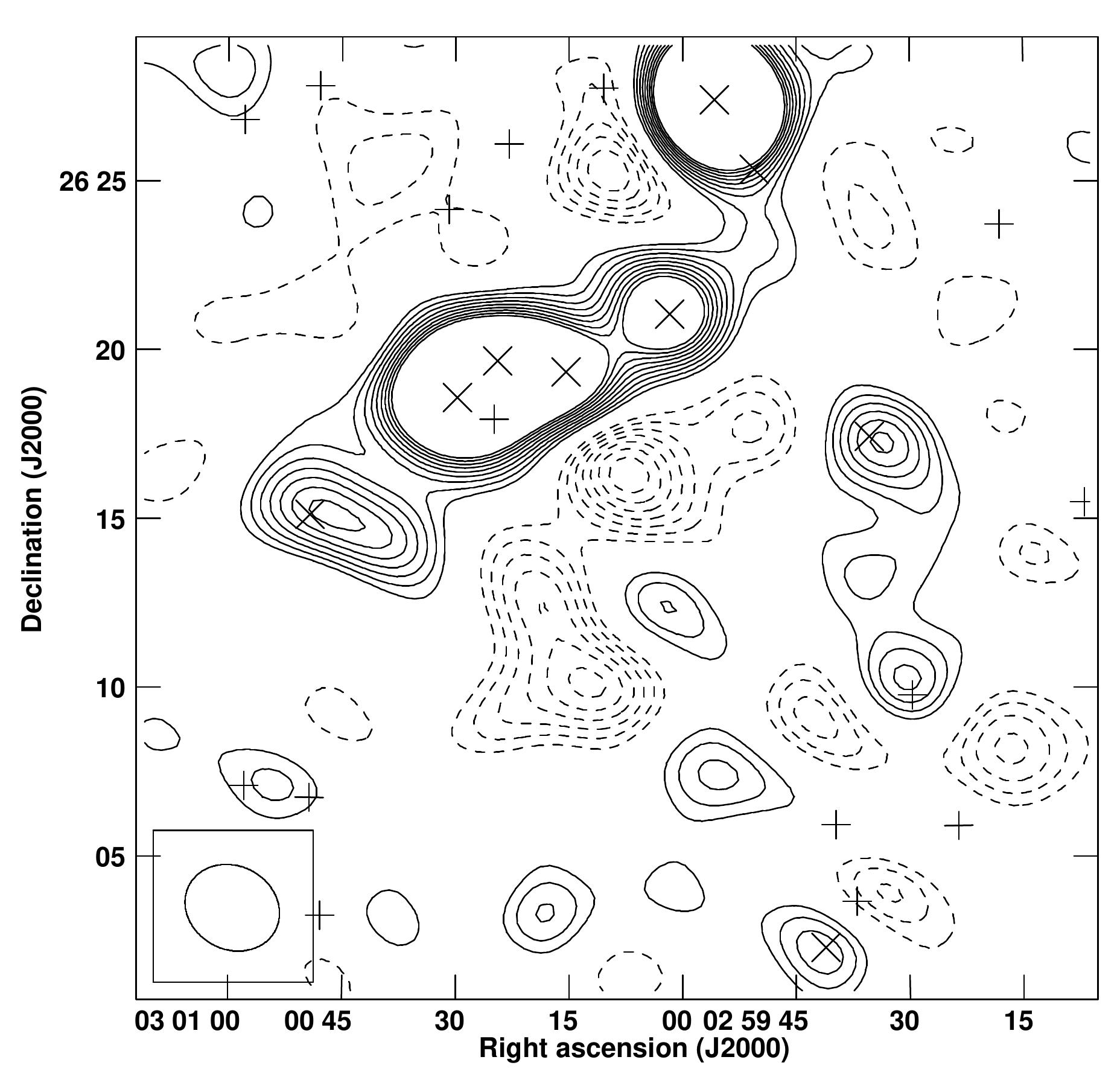}\qquad\includegraphics[width=8.0cm,clip=,angle=0.]{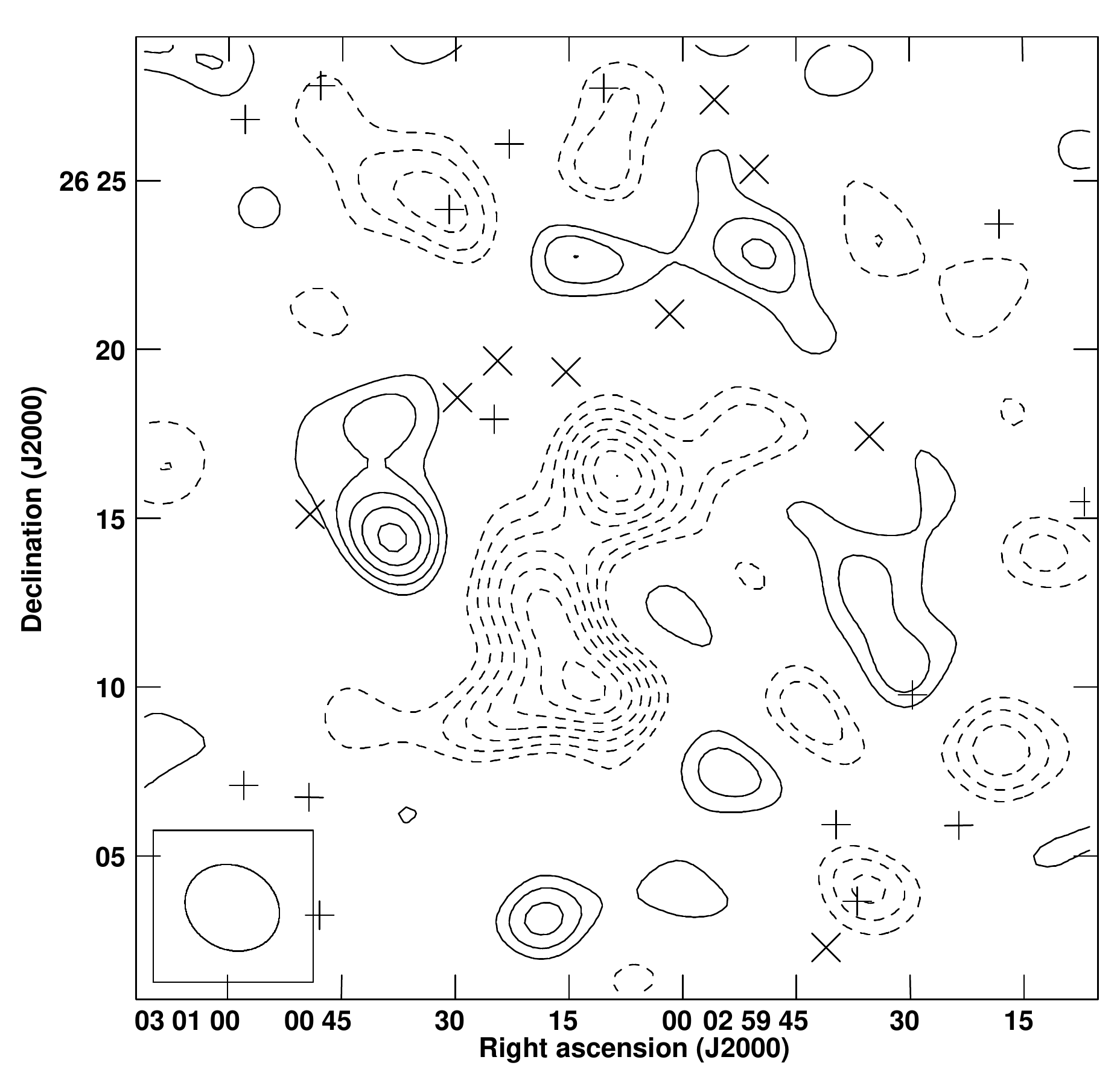}}
\caption{The AMI observation towards the cluster with contour levels that are linear from 2$\sigma_{SA}$ to 10$\sigma_{SA}$ ($\sigma_{SA}$ $=$ 65$\,\mu \rm{Jy/beam}$); positive contours are solid lines and negative contours are dashed lines. On the left is the
map before source-subtraction and on the right is the map after the detected sources have been subtracted. For both maps the visibilities are naturally weighted. The + symbols indicate
the positions of sources with flux densities less than 4$\sigma_{\rm{SA}}$, the $\times$ symbols represent sources which have a flux density greater than 4$\sigma_{\rm{SA}}$ on the SA map (see Table \ref{Radio_sources}). Cluster 1 is at J~03$^{\rm{h}}$~00$^{\rm{m}}$~14.8$^{\rm{s}}$ +26$^\circ$~10$'$~02.6$''$ and cluster 2 is at J~03$^{\rm{h}}$~00$^{\rm{m}}$~08.9$^{\rm{s}}$ +26$^\circ$~16$'$~29.1$''$. The pre source-subtraction image was \textsc{clean}ed to a depth of 3$\sigma_{SA}$ with a single \textsc{clean} box encompassing the entire image. The post source-subtraction image was \textsc{clean}ed with a circular \textsc{clean} box around the peak of each SZ decrement to a depth of 1$\sigma_{SA}$ and \textsc{clean}ing was continued to a depth of 3$\sigma_{SA}$ with a single \textsc{clean} box encompassing the entire image. Neither map has been primary beam corrected so the thermal noise is constant across the image. The ellipse at the bottom left of the maps shows the SA synthesised beam.}
\label{fig:SA_pointed}
\end{figure*}

\begin{figure*}
  \centerline{\includegraphics[width=8.0cm,clip=,angle=0.]{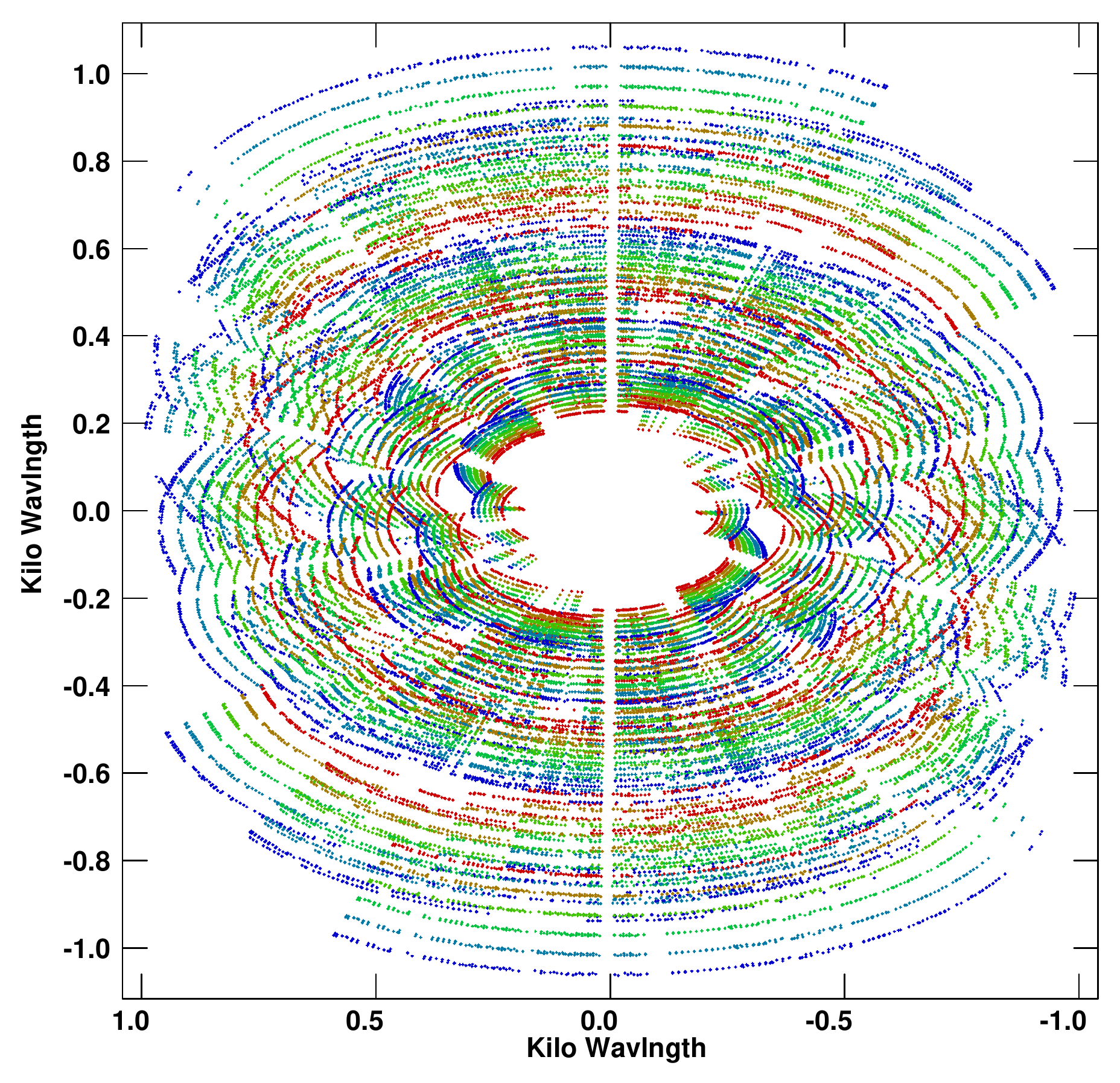}\qquad\includegraphics[width=8.0cm,clip=,angle=0.]{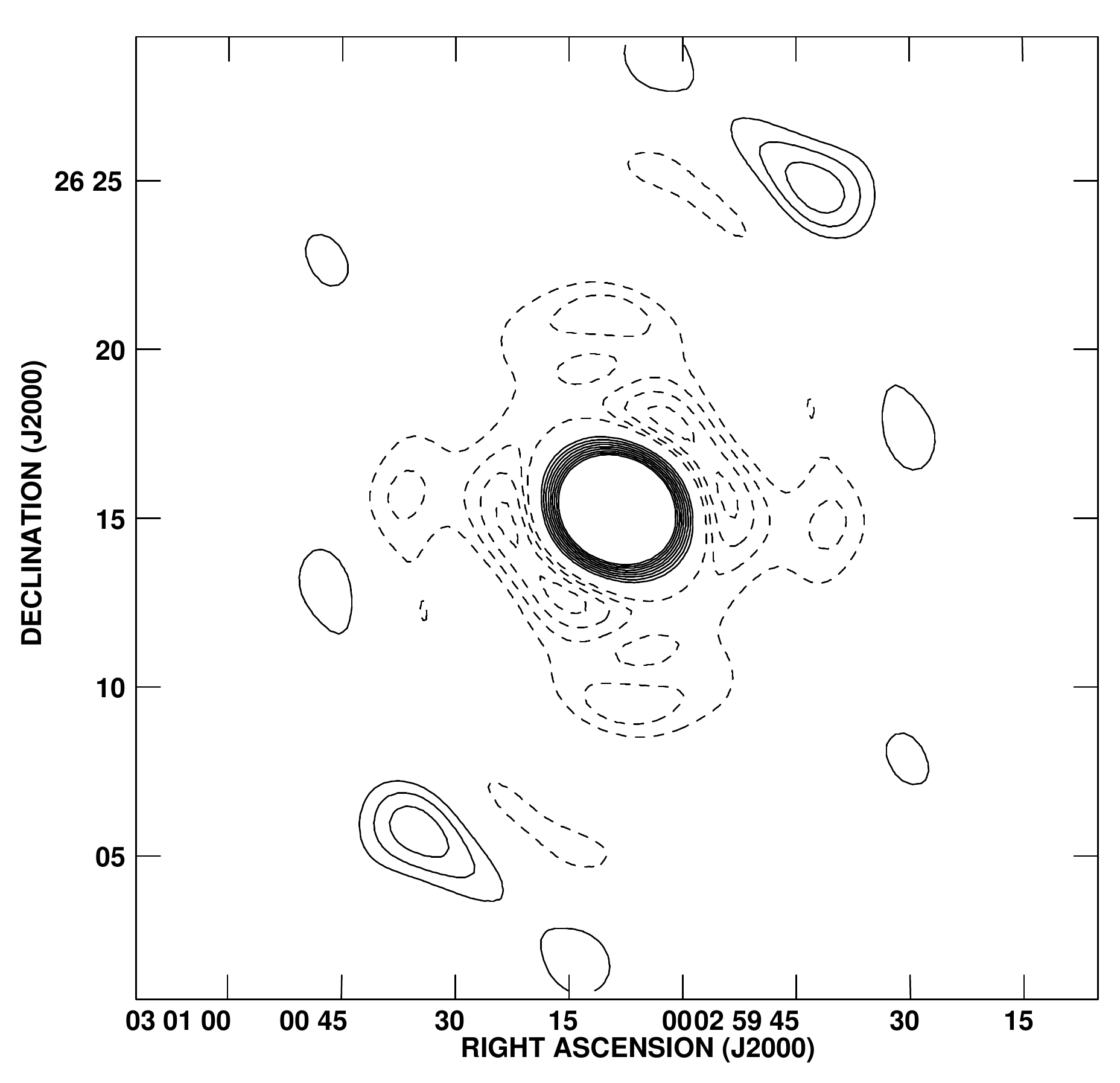}}
  \caption{Left: the $uv$-coverage of the AMI SA dataset. The different colours are for the different SA channels. Right: the naturally weighted SA synthesized beam. The synthesised beam contour levels are 6\%, 9\%, 12\% and continue to increase by 3\% up to 30\%; positive contours are solid lines and negative contours are dashed lines. \label{AMI_SA_UV}}
\end{figure*}

The first cluster detected in the survey has an extended and unusual morphology with peaks in the SZ signal at  J~03$^{\rm{h}}$~00$^{\rm{m}}$~08.9$^{\rm{s}}$ +26$^\circ$~16$'$~29.1$''$ and J~03$^{\rm{h}}$~00$^{\rm{m}}$~14.8$^{\rm{s}}$ +26$^\circ$~10$'$~02.6$''$. After the initial detection of the cluster in the survey the SA was used for deep pointed observations. In both the survey and the SA follow-up observations the SZ decrement caused by the cluster was detected with a consistent magnitude and morphology. The SA image from the pointed data is shown in Figure \ref{fig:SA_pointed}, the corresponding SA synthesized beam and $uv$-coverage are shown in Figure \ref{AMI_SA_UV} (images from Paper I). A summary of the AMI SA pointed observations towards the cluster is given in Table \ref{OBS-INFO} and these observations are described in more detail in Paper I.

\section{Follow-up Observations with CARMA}

\subsection{Instrument}

CARMA operates at either 26 -- 36\,GHz or 85 -- 116\,GHz and consists of six 10.4\,m, nine 6.1\,m, and eight 3.5\,m antennas (the 10.4\,m and 6.1\,m antennas also operate at 215 -- 265\,GHz). For these follow-up observations only the eight 3.5\,m antennas were used and we refer to these as the Sunyaev-Zel'dovich Array (SZA). Six of the eight lay in a close-packed configuration to provide 15 baselines with lengths in the range 0.4 -- 1.3\,k$\lambda$ and a resolution of $\approx$2$\arcmin$. The two outrigger antennas provide 12 longer baselines with lengths in the range 5.7 -- 9.3\,k$\lambda$ and a naturally weighted resolution of $\approx$0.3$\arcmin$. The lower resolution of the shorter baselines provides good sensitivity to the SZ effect, whereas the higher-resolution, longer baselines are insensitive to the SZ and can characterise contaminating sources.  The SZA has successfully detected the SZ effect in many clusters and has produced SZ images of clusters even at high redshift (e.g. \citealt{Muchovej_2007} and \citealt{Culverhouse_2010}). A technical summary of the SZA is given in Table \ref{tab:telescope_tech} and further details can be found in \cite{Muchovej_2007}.

\begin{table*}
\caption{A technical summary of AMI and of the SZA at 31\,GHz.}
 \label{tab:telescope_tech}
\begin{tabular}{lcccc}
\hline 
   & SA  & LA   & SZA (short) & SZA (long) \\\hline 
Antenna diameter     & 3.7\,m     & 12.8\,m    & \multicolumn{2}{|c|}{3.5\,m}   \\ 
Number of antennas & 10          & 8                & 6           & 2 (+SZA short) \\ 
Number of baselines & 45          & 28            & 15          & 12 \\ 
Baseline length (m)    & 5--20\,m     & 18--110\,m &  4--12\,m & 55--90\,m  \\ 
Baseline length (k$\lambda$) & 0.2--1.0\,k$\lambda$ &
0.9--5.8\,k$\lambda$ & 0.4--1.3\,k$\lambda$ & 5.7--9.3\,k$\lambda$   \\ 
Power primary beam FWHM  & 19.6$'$ & 5.6$'$ &   \multicolumn{2}{|c|}{11.0$'$}   \\ 
Synthesized beam FWHM &  $\approx$ 3$'$ & $\approx$ 30$''$ & $\approx$
2$'$ & $\approx$ 20$''$\\ 
Observing frequency & \multicolumn{2}{|c|}{13.5--18.0\,GHz} & \multicolumn{2}{|c|}{27--35\,GHz} \\ 
Bandwidth per sub-band & \multicolumn{2}{|c|}{4.5\,GHz}   & \multicolumn{2}{|c|}{0.5\,GHz} \\ 
Number of sub-bands & \multicolumn{2}{|c|}{1}   & \multicolumn{2}{|c|}{16} \\ 
Channels per band & \multicolumn{2}{|c|}{6}  & \multicolumn{2}{|c|}{15} \\ 
Channel bandwidth & \multicolumn{2}{|c|}{0.75\,GHz} & \multicolumn{2}{|c|}{0.03\,GHz} \\ 
Polarization measured & \multicolumn{4}{|c|}{I + Q} \\ 
Typical system temperature & \multicolumn{2}{|c|}{25K}& \multicolumn{2}{|c|}{30-60K} \\ \hline
 \end{tabular}
\end{table*}

\subsection{Observations}

\begin{figure*}
  \centerline{\includegraphics[width=8.0cm,clip=,angle=0.]{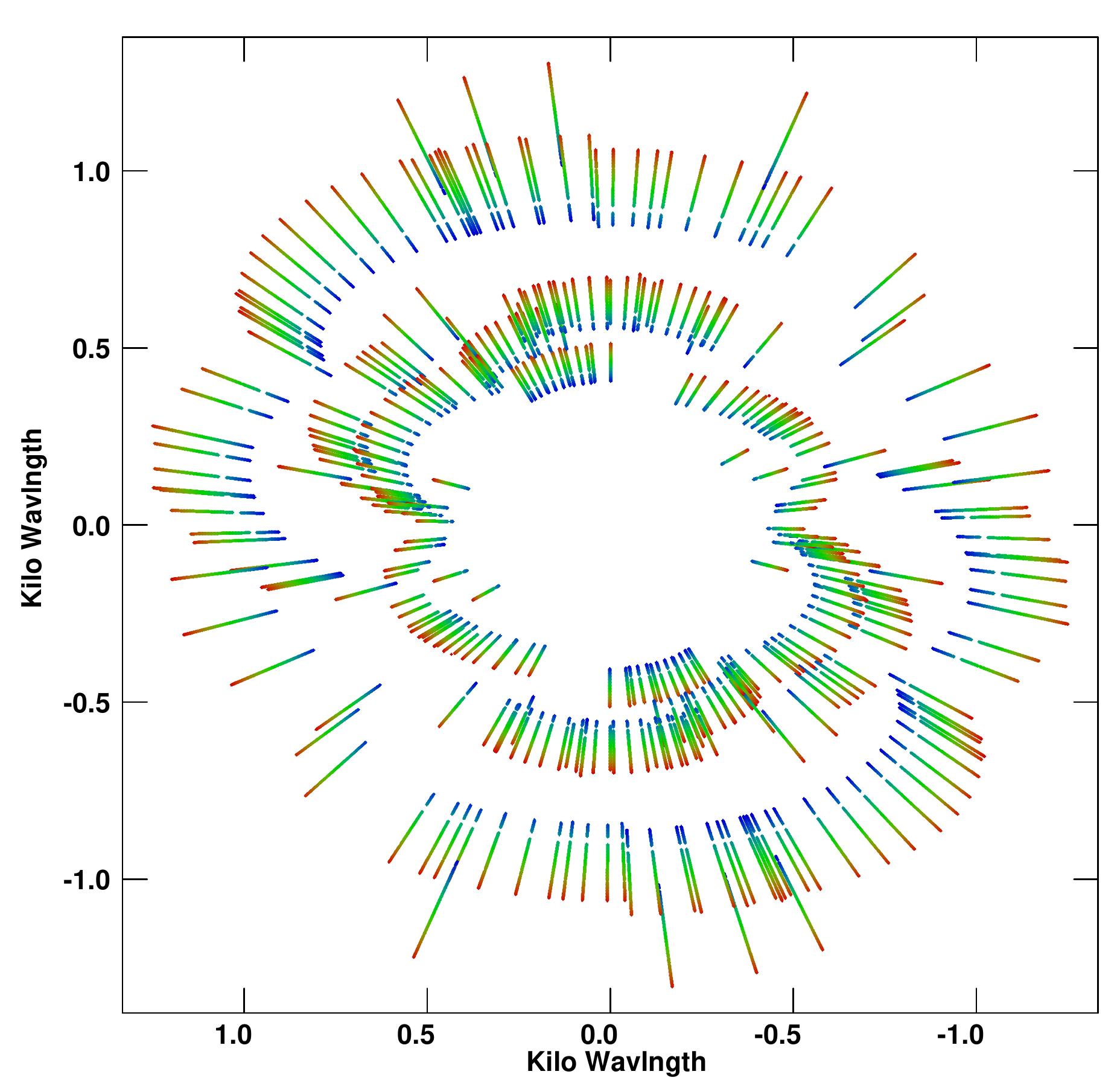}\qquad\includegraphics[width=8.0cm,,clip=,angle=0.]{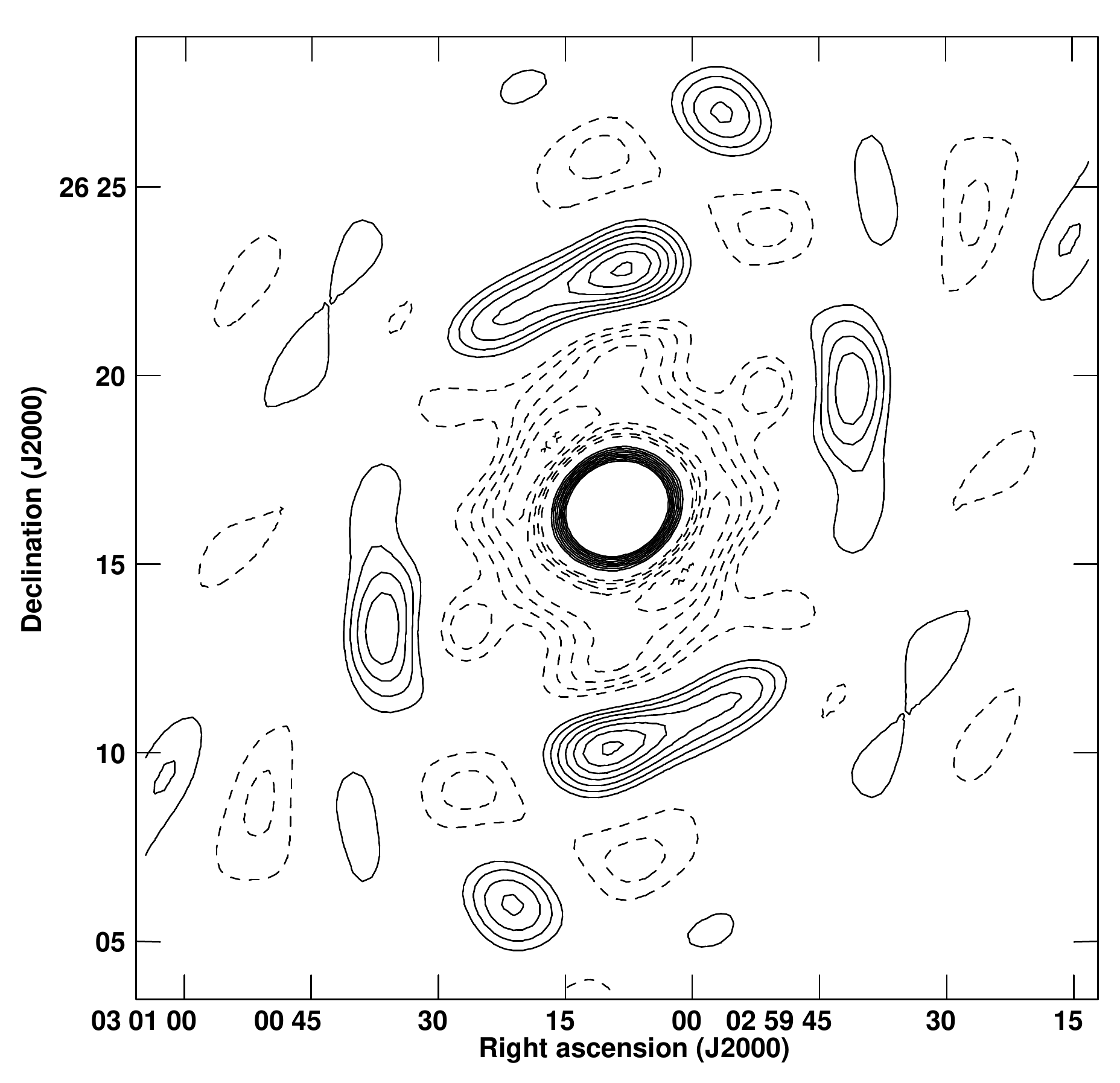}}
  \caption{Left: the $uv$-coverage of the close-packed SZA configuration dataset. The different colours are for the different SZA channels. Right: the corresponding naturally weighted synthesized beam. The contour levels are 6\%, 9\%, 12\% and continue to increase by 3\% up to 30\%; positive contours are solid lines and negative contours are dashed lines. 
  \label{CARMA_SHORT_UV}}
\end{figure*}
\begin{figure*}
  \centerline{\includegraphics[width=7.9cm,clip=,angle=0.]{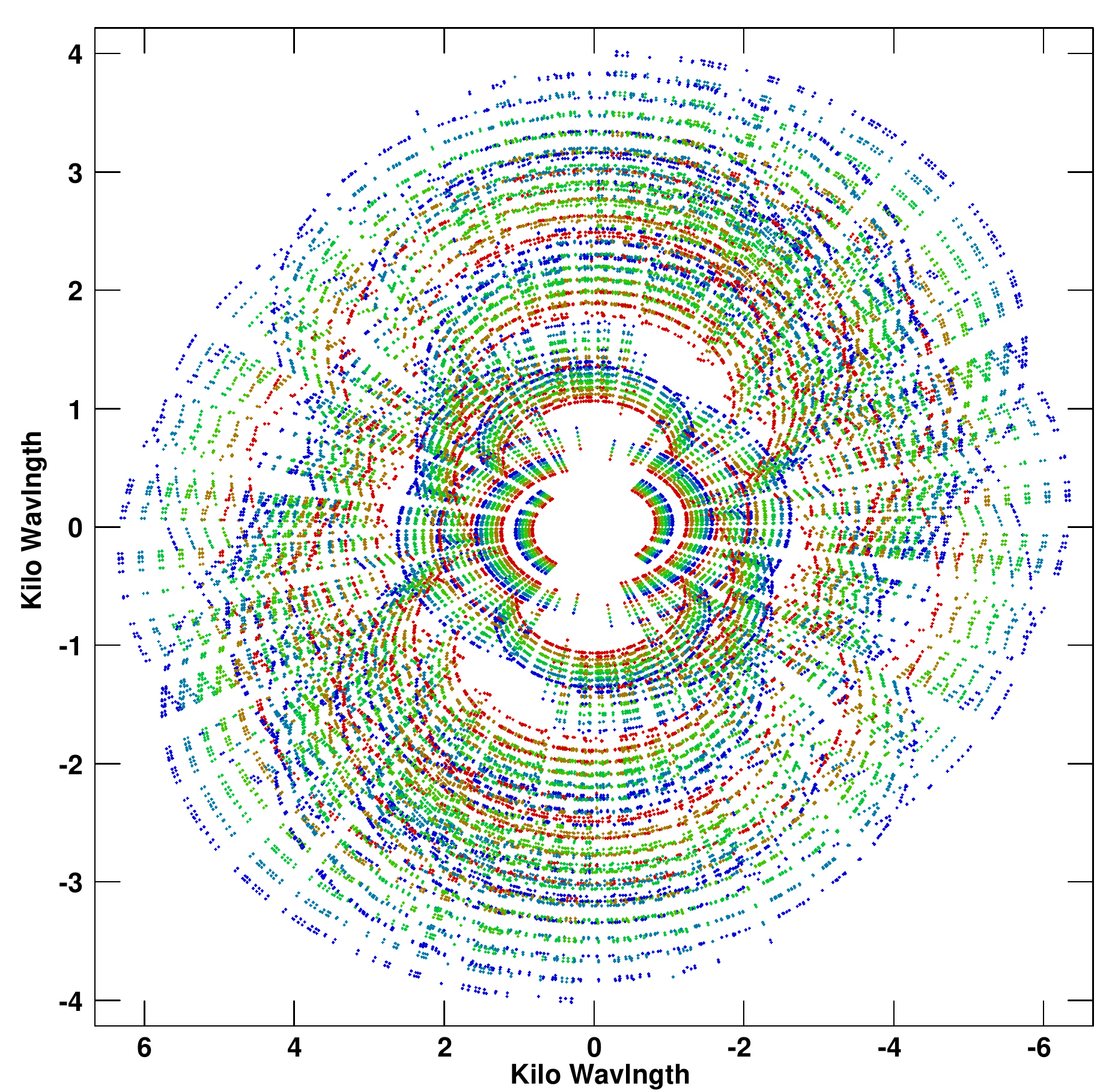}\qquad\includegraphics[width=8.0cm,,clip=,angle=0.]{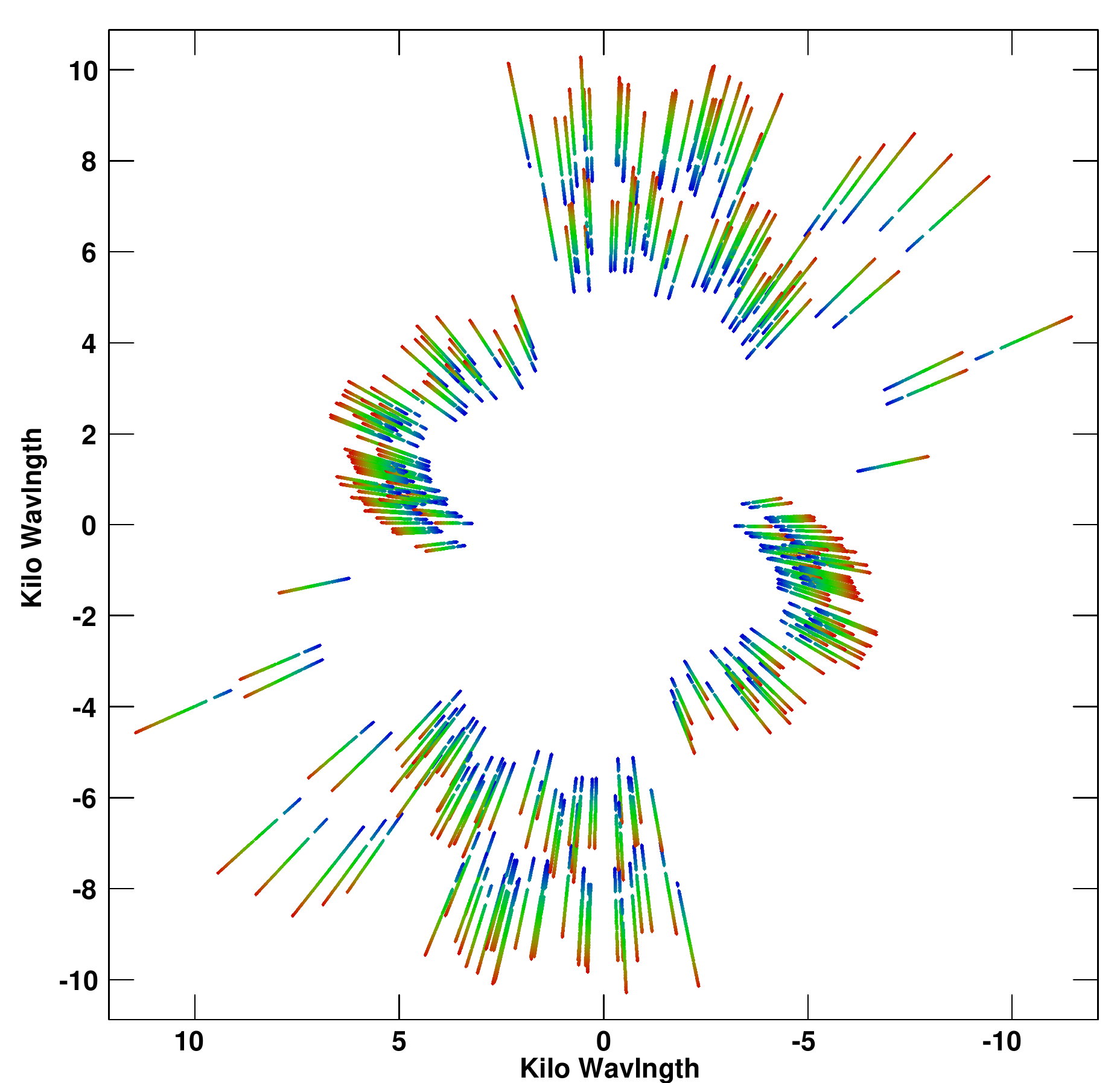}}
  \caption{Left: the $uv$-coverage for a typical pointing in the LA survey. Right: the $uv$-coverage for the SZA non-close-packed configuration. The different colours represent different channels.
  \label{CARMA_LA_UV}}
\end{figure*}

We obtained deep SZA pointed observations towards J~03$^{\rm{h}}$~00$^{\rm{m}}$~08.9$^{\rm{s}}$ +26$^\circ$~16$'$~29.1$''$; details of these observations are presented in Table \ref{OBS-INFO}. We note that the other peak in the initial results from the AMI SZ survey, J~03$^{\rm{h}}$~00$^{\rm{m}}$~14.8$^{\rm{s}}$ +26$^\circ$~10$'$~02.6$''$, lies 6.5$\arcmin$ from the SZA pointing centre at the 40\% power level of the SZA primary beam. The $uv$-coverage and corresponding synthesized beam from the close-packed configuration data are shown in Figure \ref{CARMA_SHORT_UV}.  The $uv$-coverage of the non-close-packed configuration and for comparison, the $uv$-coverage of the LA, are shown in Figure \ref{CARMA_LA_UV}.



\begin{table}
\caption{Observation summary. Note that the integration time
  corresponds to the unflagged time spent on source and the noise levels are obtained from naturally weighted visibilities.} 
 \label{OBS-INFO}
\begin{tabular}{lccc}
\hline 
   & CARMA  & AMI   \\
   & (31\,GHz) & (15.7\,GHz) \\ \hline 
Target &  \multicolumn{2}{|c|}{ J~03$^{\rm{h}}$~00$^{\rm{m}}$~08.9$^{\rm{s}}$ +26$^\circ$~16$'$~29.1$''$} \\
Integration time (hours) & 22.8  & 50.0 \\
Thermal noise ($\mu$Jy/beam) &  160 & 65  \\
Flux calibrator & Jupiter  & 3C286 \& 3C48\\
Phase (gain) calibrator & \multicolumn{2}{|c|}{0237+2848}  \\
Passband calibrator & 3C84  & 3C286 \& 3C48 \\ \hline
 \end{tabular}
\end{table}

\subsection{Data Reduction and Imaging}

\begin{figure*}
\centerline{\includegraphics[width= 8.0cm,clip=,angle=0.]{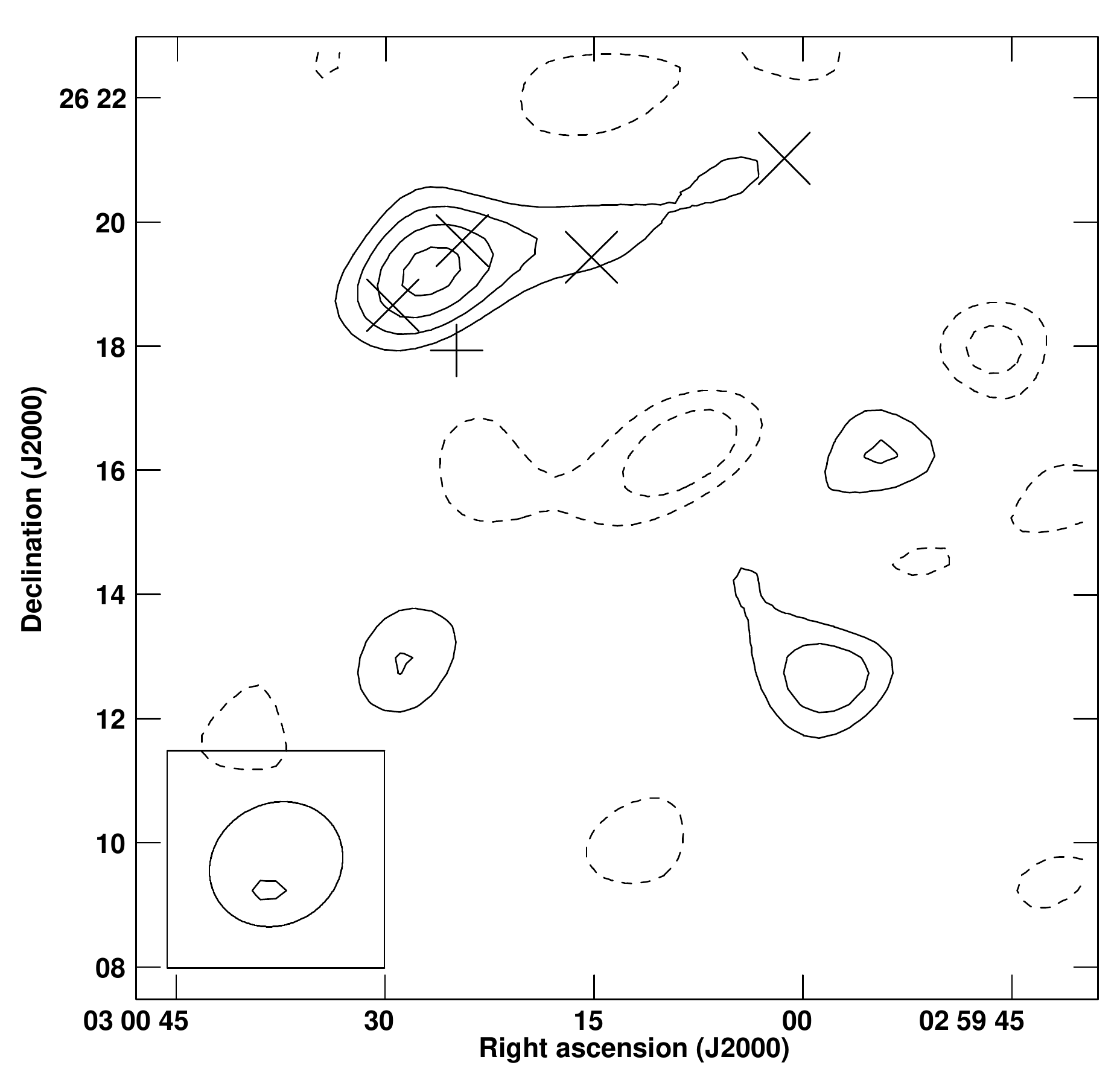}\qquad\includegraphics[width=8.0cm,clip=,angle=0.]{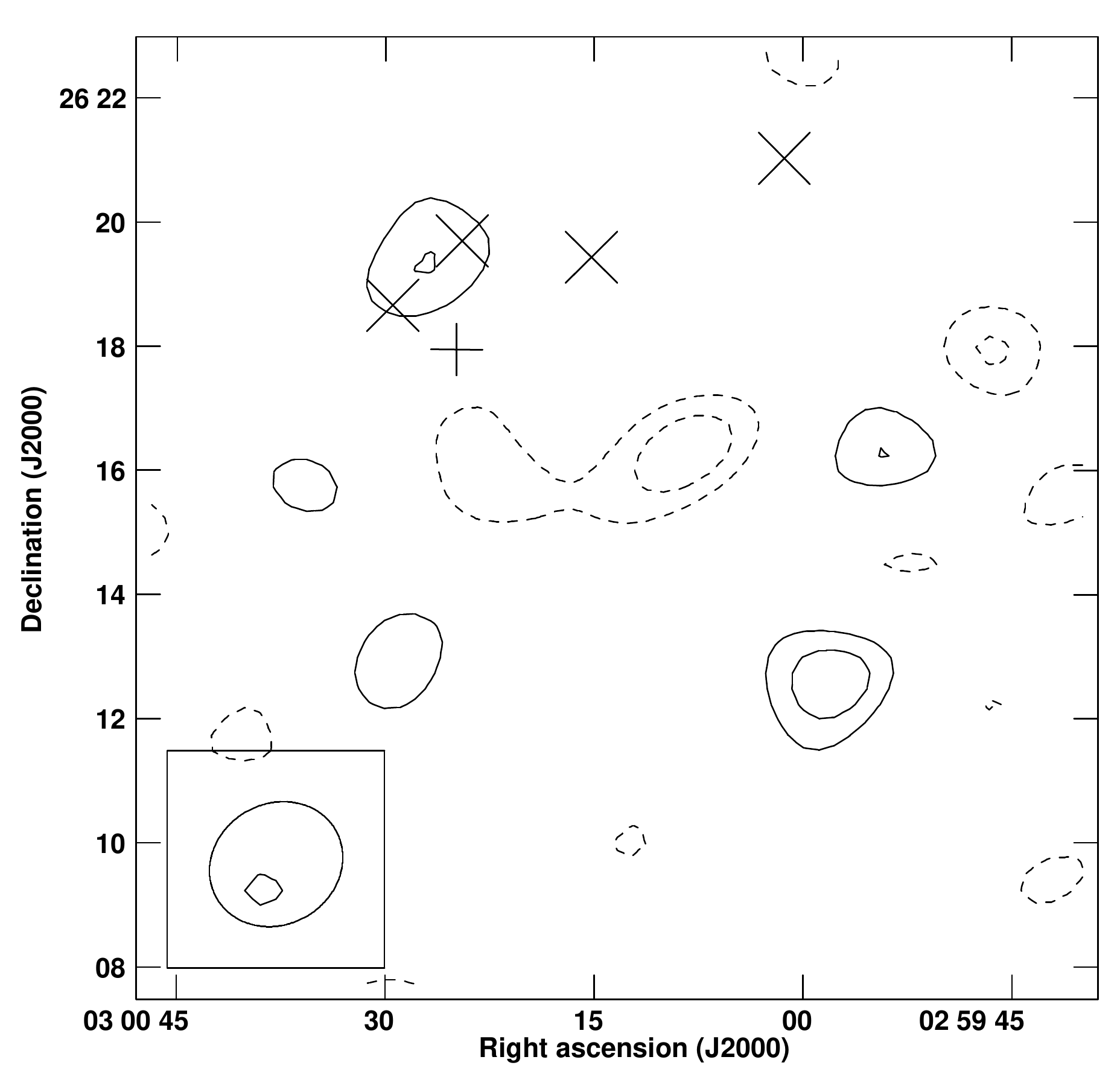}}
\caption{The SZA observation towards the cluster with contour levels that are linear from 2$\sigma_{SZA}$ to 10$\sigma_{SZA}$ ($\sigma_{SZA}$ $=$ 160$\,\mu \rm{Jy/beam}$); positive contours are solid lines and negative contours are dashed lines. On the left is the map before source-subtraction and on the right is the map after the detected sources have been subtracted. The + symbols indicate sources with a low flux that have been subtracted but not modelled and the $\times$ symbols represent sources that have been subtracted using the modelled parameters (see Table \ref{Radio_sources}). The SZA visibilities were naturally weighted but only baselines shorter than 2k$\lambda$ were used. The pre source-subtraction image was \textsc{clean}ed to a depth of 3$\sigma_{SZA}$ with a single \textsc{clean} box encompassing the entire image. The post source-subtraction image was \textsc{clean}ed with a circular \textsc{clean} box around the SZ decrement to a depth of 1$\sigma_{SZA}$ and \textsc{clean}ing was continued to a depth of 3$\sigma_{SZA}$ with a single \textsc{clean} box encompassing the entire image. Neither map has been primary beam corrected so the thermal noise is constant across the image. The ellipse at the bottom left of the maps shows the SZA synthesised beam.
\label{fig:SZA_pointed}}
\end{figure*}

\begin{figure*}
\centerline{\includegraphics[width= 8.0cm,clip=,angle=0.]{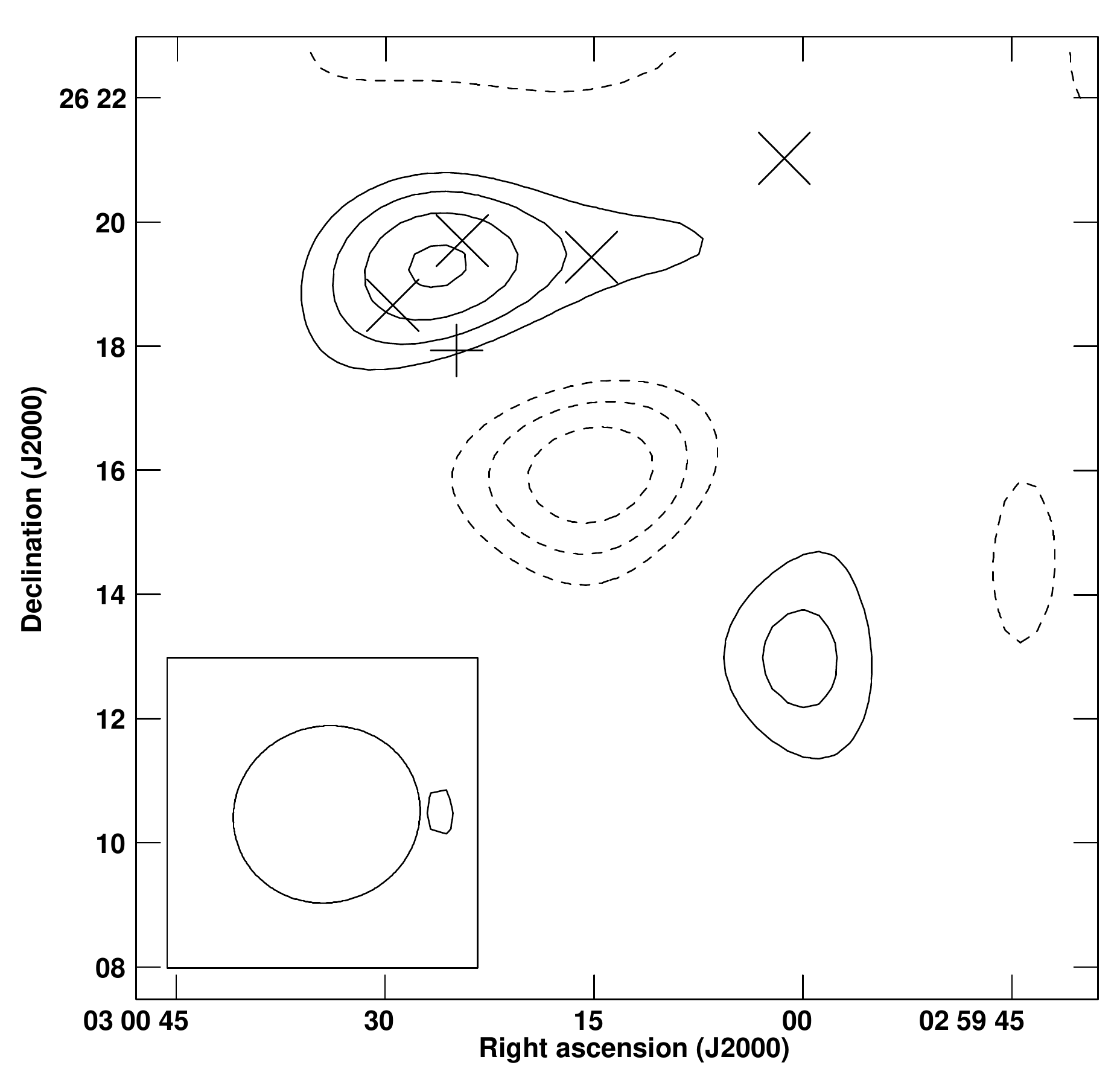}\qquad\includegraphics[width=8.0cm,clip=,angle=0.]{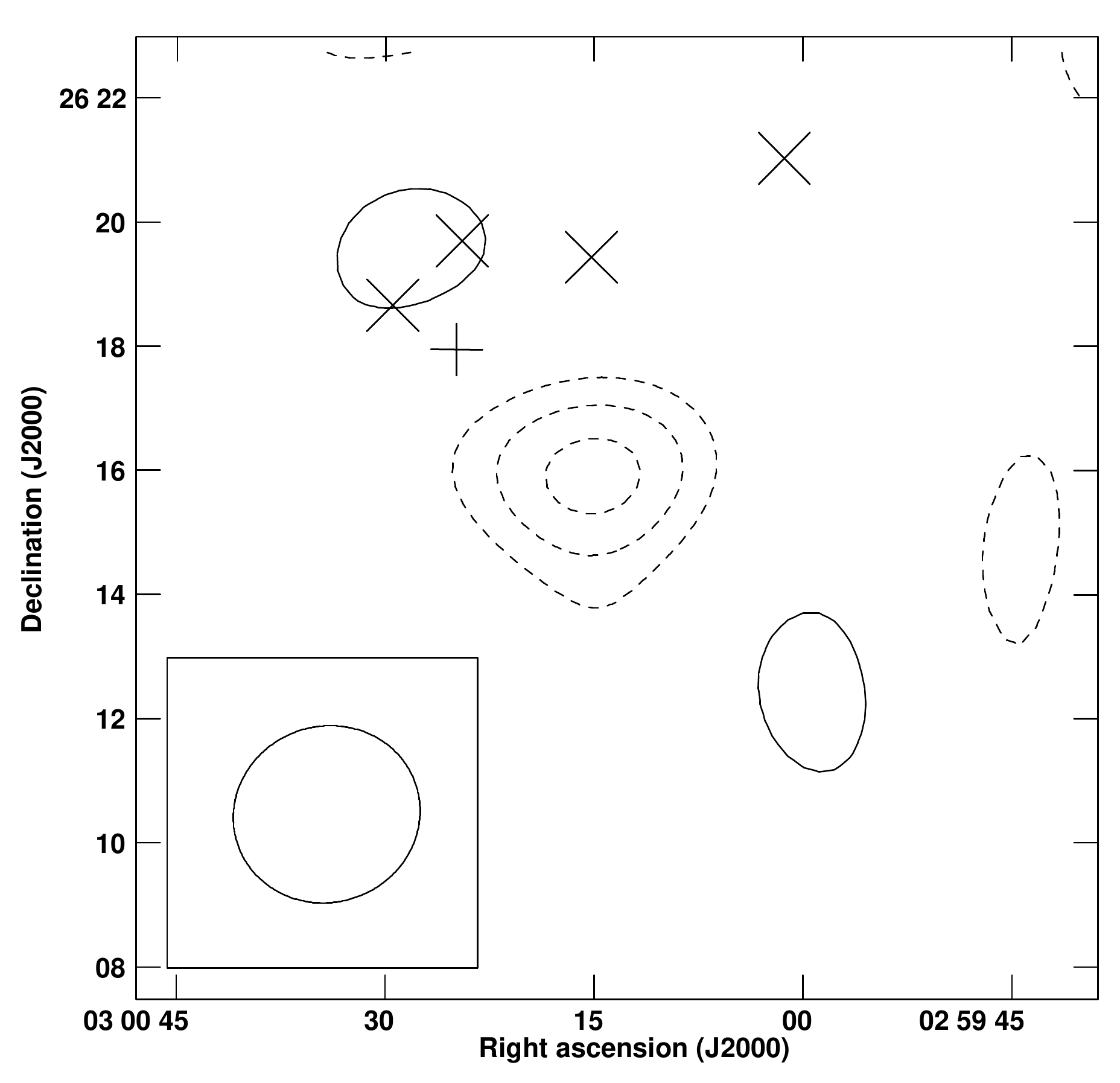}}
\caption{The same as Figure \ref{fig:SZA_pointed} but the SZA visibilities were weighted with a Gaussian function during the imaging procedure in \textsc{aips} and the distance to the 30\% point of the Gaussian was set to 0.5k$\lambda$ -- this provides a resolution of $\approx 3\arcmin$. The contour levels are linear from 2$\sigma_{SZA-TAPER}$ to 10$\sigma_{SZA-TAPER}$ ($\sigma_{SZA-TAPER}$=217\,$\mu$Jy/beam);  positive contours are solid lines and negative contours are dashed lines.
\label{fig:SZA_pointed_taper}}
\end{figure*}

The SZA data was calibrated with the \textsc{miriad} package
(\citealt{Sault_1995}). In this software the data were flagged for interference, bad
antennas and sub-bands. Further manual excision of bad data was performed in \textsc{casa}\footnote{http://casa.nrao.edu/} and imaging was done in \textsc{aips}. To enhance low surface brightness structures in the SZA SZ imaging procedure we did not use
the visibilities from the outrigger antennas.  An image of the naturally weighted SZA data from the short baselines is presented in Figure \ref{fig:SZA_pointed}. An image where low surface brightness structures were further enhanced by applying a Gaussian weighting function to the visibilities is presented in Figure \ref{fig:SZA_pointed_taper}.

To search for point sources in the SZA data, the visibilities from the 12 baselines provided by the outrigger antennas were imaged together with the data from shorter baselines. The data were imaged with uniform weighting to produce a high-resolution ($\approx 20\arcsec$) image with a sensitivity of 180\,$\mu$Jy/beam. Deconvolution was not performed because the $uv$-coverage is sparse and the sidelobes are high (first negative sidelobe $\approx25$\% and the first positive sidelobe $\approx50$\%). The image is only suitable for the detection of bright sources; there however are no bright sources in this field and we are unable to detect any sources with fluxes greater than 4 times the thermal noise in this high-resolution SZA image.


\section{Analysis}

To analyse the AMI data and SZA data, and both datasets together, we use a Bayesian analysis (\citealt{MH_MCADAM}, \citealt{MARSH_MCADAM} and \citealt{FF_MCADAM}). Given a parameterisation for the galaxy cluster and radio sources, together with a prior on each parameter probability distribution, our analysis uses the {\sc MultiNest} sampler (\citealt{MULTINEST_1} and \citealt{MULTINEST_2}) to efficiently calculate the probability distributions for all parameters in the cluster and source parameterisations.  Furthermore {\sc MultiNest} outputs the Bayesian evidence of the model which can be used for model comparison. 

\subsection{Cluster modelling}

Similarly to Paper I, we perform a Bayesian analysis of our data with a physical cluster model and a phenomenological cluster model. Both analysis techniques parameterise the observed SZ decrement and quantify the significance of detection accounting for the statistics of the primordial CMB, radio sources and receiver noise.

\subsubsection{Physical model}

 In our physical $\beta$-model we characterise the cluster with the parameters $x_{c}$, $y_{c}$, $\phi$, $f$, $\beta$, $r_{c}$, $M_{T,200}$, and $z$, here $x_{c}$ and $y_{c}$ give the cluster position, $\phi$ is the orientation angle measured from N
through E, $f$ is the ratio of the lengths of the semi-minor to semi-major axes, $\beta$ and $r_c$ (the core radius) describe the cluster gas density, $\rho_{g}$, according to \cite{BETA_1}, $M_{T,200}$ is the cluster total mass within a radius $r_{200}$ and $z$ is the cluster redshift.  The priors on the physical $\beta$-model cluster parameters are the same as used in Paper I and are given in Table \ref{MC_PRIORS}. For comparison with results from the physical $\beta$-model analysis, we  analyse the joint AMI \& SZA data with the parametric model described in \citealt{Olamaie_2012} and the shape of the pressure profile fixed according to \cite{Arnaud_2010}. In this analysis we use the same priors on $M_{T,200}$ and $z$ as those used in the analysis with the $\beta$-model. 

{With the assumption that the cluster is virialized, isothermal, spherical and that all kinetic energy in it is plasma internal energy, the $M_{\rm{T,200}}$ that we derive in our physical model analysis is used to ascertain the mean gas temperature, $T_{e,200}$, within $r_{200}$
\begin{align} 
 K_BT_{e,200} = \frac{G\mu M_{T,200}}{2r_{200}},
 \label{eqn:M-T-relation}
 \end{align}
where $\mu$ is the mass per particle ($\mu \approx 0.6$ times the mass of a proton) and $r_{200}$ is defined as the radius inside which the mean total density is 200 times the critical density of the Universe at the cluster redshift, $\rm{\rho_{crit,z}}$.

To quantify the significance of a detection, we define a cluster to have a mass $M_{T,lim} < M_{T,200} < M_{T,max}$  and we quote the ratio of Bayesian evidences, $R$, normalised by the expected number of clusters, $\mu_S$, within a search region $S$.

\begin{equation}
R \approx \frac{Z_{1}(S)\mu_S}{Z_{0}},
\label{eqn:R5}
\end{equation}
where $Z_{1}(S)$ is the evidence associated with a cluster and $Z_{0}$ is the `null' evidence.  \cite{Jeffreys_1961} provides an interpretive scale for
the $R$ value, as do revised scales  such as \cite{Gordon_2007}. 
The $R$ value can be turned into a Bayesian probability, $p$, that a cluster has been detected with $M_{T,lim} < M_{T,200} < M_{T,max}$,

\begin{equation}
p = \frac{R}{1+R}.
\label{eqn:p}
\end{equation}

For our analysis of the AMI, SA and joint AMI \& SZA data we set $M_{T,lim}$ = $2.0 \times$ $10^{14}$ and $M_{T,max}$ = 5 $\times$ $10^{15}$; the same values were used in Paper I. We use the same $M_{T,lim}$ on each dataset even though the noises of the SA data and the SZA data are different. Given that the intensity of the SZ effect is expected to be a factor of 3.6 times higher at 31\,GHz than at 15.7\,GHz and the noise is only a factor of 2.5 lower on the SA we consider this $M_{T,lim}$ to be conservative for the analysis of the SZA data or the joint AMI \& SZA data. 

For the analysis of the SZA data we chose to use two different values for $\mu_S$: a) if we consider the SA detection is genuine then we know a priori that the cluster is present and $\mu_S$ is unity; b) if we assume that \cite{Jenkins_2001} accurately predicts the detectable (0.2 $<z<$2.0 and M$\geq 2\times 10^{14}$) number of clusters within the search area then $\mu_S$=0.085. Although there are many more recent attempts to estimate the cluster number counts (e.g. \citealt{Evrard_2002}, \citealt{Sheth2002}, \citealt{White2002}, \citealt{Reed2003}, \citealt{Heitmann2006}, \citealt{Warren2006}, \citealt{Reed2007}, \citealt{Lukic2007}, \citealt{Tinker2008}, \citealt{Boylan-Kolchin2009}, \citealt{Crocce2010} and \citealt{Bhattacharya2011}), the number of clusters as a function of mass and redshift predicted by these more recent estimates within our $M_{T,200}$ and $z$ range of interest is similar to that which we have used. 

For analysis a) we make use of our prior knowledge of the cluster position and we use a Gaussian prior centred on the AMI derived coordinates to describe the cluster position. For analysis b) we instead use a uniform prior that encompasses the telescope primary beam to describe the cluster position.

\begin{table*}
\caption{Priors used for the Bayesian analysis assuming a physical
  cluster model. The mass prior extends down to $M_{T,200} = 1.5 \times10^{14}M_{\odot}$ but we note that $M_{T,lim}=2.0\times10^{14}M_{\odot}$. To blindly search for cluster signatures we used the blind cluster position prior but if we make use of our prior knowledge of the cluster then we can use the known cluster position prior.}
 \label{MC_PRIORS}
\begin{tabular}{lcc}
 \hline
Parameter & Prior \\  \hline
Redshift ($z$) & Joint prior with $M_{T}$ between 0.2 and 2.0 (\citealt{Jenkins_2001})\\ 
Core radius ($r_{c}\rm/{kpc}$) & Uniform between 10 and 1000 \\ 
Beta   ($\beta$)  & Uniform between 0.3 and 2.5 \\ 
Mass ($M_{T,200}/M_{\odot}$) & Joint prior with $z$ between 1.5 $\times$ $10^{14}$ and 5 $\times$ $10^{15}$ (\citealt{Jenkins_2001})\\ 
Gas fraction ($f_{g}$) & Delta-function prior at 0.11  (\citealt{WMAP_FGAS_VALUE}) for $Z_{1}(S)$ or at 0.0 for $Z_{0}$ \\ 
Blind cluster position ($\bf x_{c}$) & AMI -- Uniform search region between $-500\arcsec$ and $500\arcsec$\\
                                              & SZA and AMI \& SZA -- Uniform search region between $-250\arcsec$ and $250\arcsec$\\
Known cluster position ($\bf x_{c}$) & Gaussian search region centred on the cluster location with a $\sigma$ of 60$\arcsec$ \\
Orientation angle ($\phi/\deg$) & Uniform between 0 and 180 \\ 
Ratio of the length of semi-minor to semi-major axes ($f$) & Uniform between 0.5 and 1.0 \\ \hline
\end{tabular}
\end{table*}

\subsubsection{Phenomenological model}

At the location of the SZ decrement, which is obtained from the physical cluster analysis, we fit an elliptical $\beta$ profile to characterise the shape and magnitude of the temperature decrement with the cluster position and ellipticity parameters together with $\theta_c$, $\beta$ and $\Delta T_0$ (see Paper I). The assumed priors for these parameters are given in Table \ref{MC_PHEN_PRIORS}.

\begin{table}
\caption{Priors used for the Bayesian analysis assuming a phenomenological 
  cluster model. }
 \label{MC_PHEN_PRIORS}
\begin{tabular}{lcc}
 \hline
Parameter & Prior \\  \hline
$\Delta T_{0}$ & Uniform between $\pm$ 3000$\mu$K \\
$\theta_c$ & Uniform between 20 and 500$\arcsec$ \\
$\beta$ & Uniform between 0.4 and 2.5 \\ \hline
\end{tabular}
\end{table}

\subsection{Source modelling}

In our analysis sources are characterised with the parameters $x_{s},y_{s},S_{0}$ and $\alpha$, where  $x_{s}$ and $y_{s}$ are the source coordinates, $S_{0}$ is the source flux at frequency $\nu_0$ and $\alpha$ is the source spectral index ($S\propto \nu^{-\alpha}$).

The LA was used to determine the source environment at 15.7\,GHz within the SA survey regions. In the survey field containing this cluster the thermal noise of the LA observations was typically 50\,$\mu$Jy/beam, but towards the cluster deeper observations were obtained to reach a 30\,$\mu$Jy/beam thermal noise level. In the one square-degree survey field containing this cluster 203 sources were detected by the LA with flux densities greater than 4 times the LA thermal noise level, $\sigma_{LA}$. Most of these sources lie far away from this target and Paper I presented a list of all nine sources which have a 15.7\,GHz flux density greater than 4 times the SA thermal noise of the pointed observation ($\sigma_{SA}$; 65\,$\mu$Jy/beam at the pointing centre). 

For the analysis of our SA observations we use the source positions, spectral indexes, flux densities and the mean weighted frequency from the LA catalogue. We put delta priors on all faint sources (less than 4$\sigma_{SA}$) and allow our analysis to model all source parameters for the nine brighter sources, see Table \ref{Radio_sources} for the priors on the modelled sources.

Precise priors on the source environment at 31\,GHz would greatly assist with the analysis of the SZA data. Extrapolating source fluxes from low-frequency measurements (e.g. NVSS;  \citealt{Condon_1998}) combined with our 15.7\,GHz measurements will not provide reliable estimates of the 31\,GHz flux (see e.g. \citealt{Gawronski_2010}). Instead we attempt to estimate the flux levels from high-resolution SZA images in which the extended SZ effect is resolved out. We map the SZA data from all baselines using uniform weighting and no deconvolution to produce a high-resolution image ($\approx$20$\arcsec$) with a thermal noise level of 180\,$\mu$Jy/beam. In this image we detect no sources with fluxes $\geq 4.0$ times the thermal noise, but 
for the four sources within the half-power point of the SZA primary beam we extract from this map SZA flux estimates; these are all $\geq 2$ times the thermal noise and are used as the Bayesian priors. 
All sources further than 50\% of the way down the primary beam are undetected in our high-resolution SZA data and for these sources we use the LA flux measurement with a 40\% error and the LA derived spectral index as Bayesian priors.

For the joint SA and SZA analysis we use the same source priors as used in the SZA only analysis.

\begin{table}
\caption{Parameters for the nine sources modelled in our analyses of the AMI, the SZA and the joint AMI \& SZA data. At 15.7\,GHz these have flux densities exceeding  4$\sigma_{SA}$ ($\sigma_{SA}=65\,\mu$Jy/beam at the pointing centre) but at 31\,GHz the flux densities do not exceed 4$\sigma_{SZA}$ ($\sigma_{SZA}=160\,\mu$Jy/beam at the pointing centre). The positions and mean frequencies are from LA observations, whereas the AMI and SZA flux densities are the values obtained from our Bayesian analysis of the SA or SZA data respectively.}
 \label{Radio_sources}
\begin{tabular}{lcccc}
 \hline
Right ascension & Declination  &  AMI flux       & SZA flux         \\
                          &                      & density         & density         \\ 
(J2000)               & (J2000)          & (mJy)             &      (mJy) \\ \hline
02:59:55.16 & +26:27:26.25 & 8.58$\pm$0.17 & 3.17$\pm$1.80 \\ 
03:00:49.28 & +26:15:05.07 & 0.87$\pm$0.09 & 0.48$\pm$0.23 \\
03:00:24.53 & +26:19:40.83 & 1.41$\pm$0.11 & 0.46$\pm$0.33 \\
03:00:01.33 & +26:21:01.55 & 1.50$\pm$0.08 & 0.28$\pm$0.52 \\
03:00:29.46 & +26:18:39.95 & 2.23$\pm$0.09 & 0.43$\pm$0.36 \\
02:59:41.05 & +26:02:20.41 & 1.65$\pm$0.22 & 0.62$\pm$0.59 \\
02:59:50.35 & +26:25:22.37 & 0.54$\pm$0.12 & 0.38$\pm$0.24 \\
03:00:15.23 & +26:19:25.56 & 1.23$\pm$0.08 & 0.36$\pm$0.31 \\
02:59:35.43 & +26:17:26.77 & 0.69$\pm$0.08 & 0.50$\pm$0.21 \\ \hline
 \end{tabular}
\end{table}

\section{Results and discussion}

We have presented source-subtracted images of the SA and SZA data but, due to the finite sampling of the $uv$-plane, an accurate analysis of inteferometric images of extended objects is difficult (Figure \ref{fig:AMI+SZA-map}). Simply quantifying the significance of the SZ decrement on the image does not account for CMB contamination, possible degeneracies between source and cluster parameters (which is not the case for these data), uncertainties in the source fluxes of subtracted sources and general imaging problems including the \textsc{clean} bias which is most severe for faint features such as SZ decrements. Nevertheless, the SZ flux density is expected to be a factor of 3.6 times greater at  31\,GHz than at 15.7\,GHz but in our images the peak decrement at 31\,GHz (620\,$\mu$Jy; 3.9$\sigma_{SZA}$) is only 117\% of that measured at 15.7\,GHz (530\,$\mu$Jy; 8$\sigma_{SA}$). We are able to increase the magnitude of the peak decrement to 960\,$\mu$Jy (4.0$\sigma_{SZA-TAPER}$) on the SZA image by weighting the visibilities with a Gaussian taper in the $uv$-plane but the decrement is still significantly weaker than expected. The lower than expected magnitude of the SZA signal can be explained if the SZ signal is partially resolved out by the SZA. The analysis of the SA data reveals that the angular extent of the cluster is large, from this analysis we obtain $r_{200}=1.64 \pm 0.13 $Mpc 
 which at the mean derived redshift of 0.68 (the redshift is constrained by the \citealt{Jenkins_2001} prior) corresponds to an angular radius of 3.9$\arcmin$. With the data naturally weighted both the SA and the SZA will resolve an object of this size but the SA is sensitive to larger angular scales than the SZA (see Figure \ref{fig:sa-sza-resolution-sensitivity}) and thus we expect that the SZA is resolving out more of the signal than the SA.  To explore this further we could replicate the SZA $uv$-coverage with the SA data, but this operation results in a large data loss which consequently raises the thermal noise significantly and makes any comparison between the AMI and SZA data difficult. Alternatively, using the mean values of the derived parameters from the analysis of the AMI data, realistic simulations of AMI and SZA datasets could be created and analysed in an attempt to reproduce the discrepancy. However with the AMI and SZA joint analysis pipeline now in place, this discrepancy will be best investigated in a study we are presently undertaking that involves a thorough analysis of both simulations and real AMI and SZA observations of a sample of well known clusters that spans a wide range of extensions and morphologies. Ideal systems for this study are those which are at a known redshift, have been studied at X-ray wavelengths, and have total SZ signal measurements. 

\begin{figure}
  \includegraphics[width=7.5cm,clip=,angle=0.]{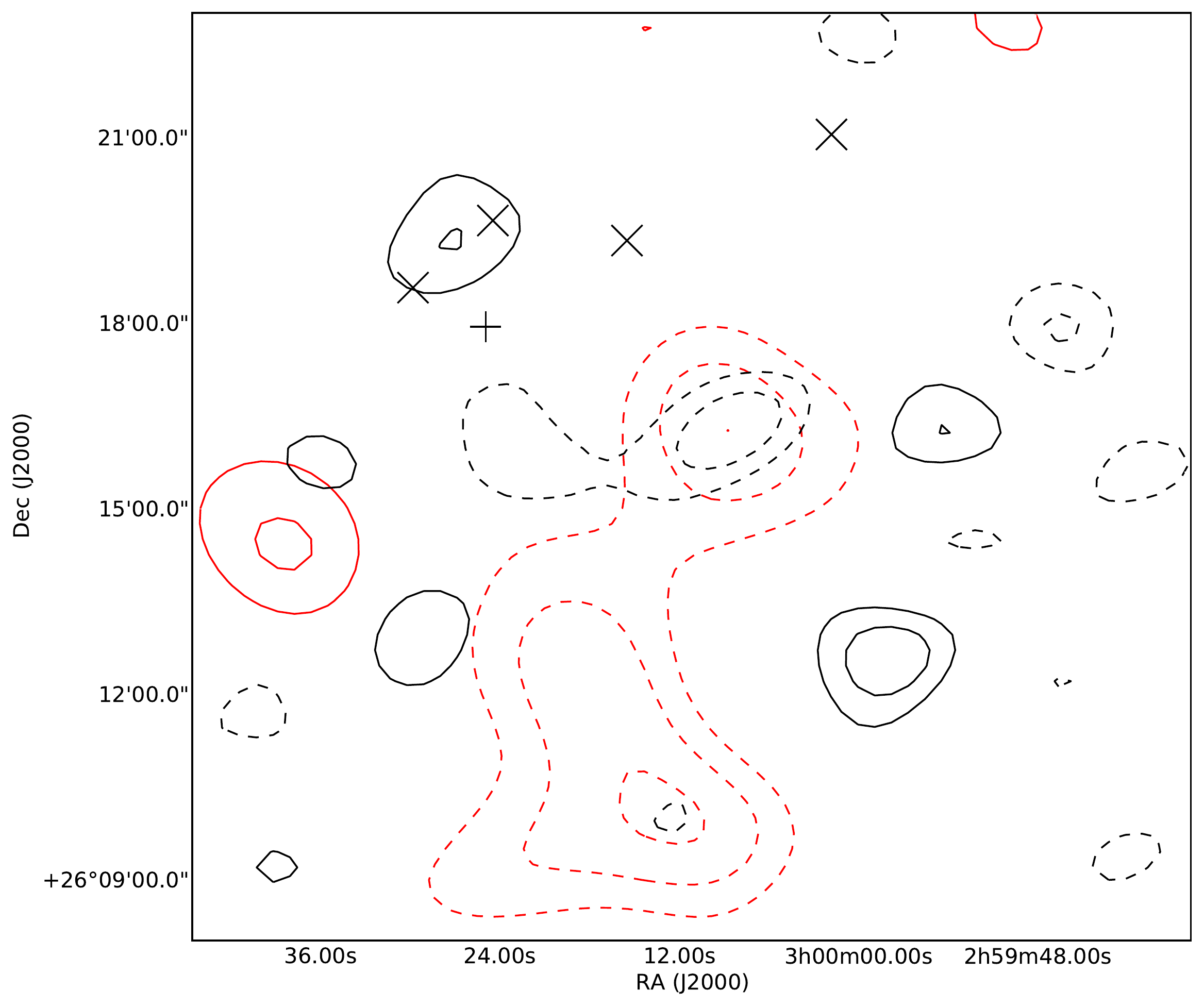}
  \caption{An overlay of the source-subtracted SA (Figure \ref{fig:SA_pointed}) and SZA (Figures \ref{fig:SZA_pointed} and \ref{fig:SZA_pointed_taper}) images. The SZA (black; $\sigma_{SZA}=160\,\mu$Jy/beam) contour levels are linear from 2$\sigma$ to 10$\sigma$, the SA (red; $\sigma_{SA}=65\,\mu$Jy/beam) contour levels range from 4$\sigma$ to 10$\sigma$ with 2$\sigma$ intervals. Dashed contours are negative and solid contours are positive.}
  \label{fig:AMI+SZA-map}
\end{figure}

\begin{figure}
  \includegraphics[width=8.5cm,clip=,angle=0.]{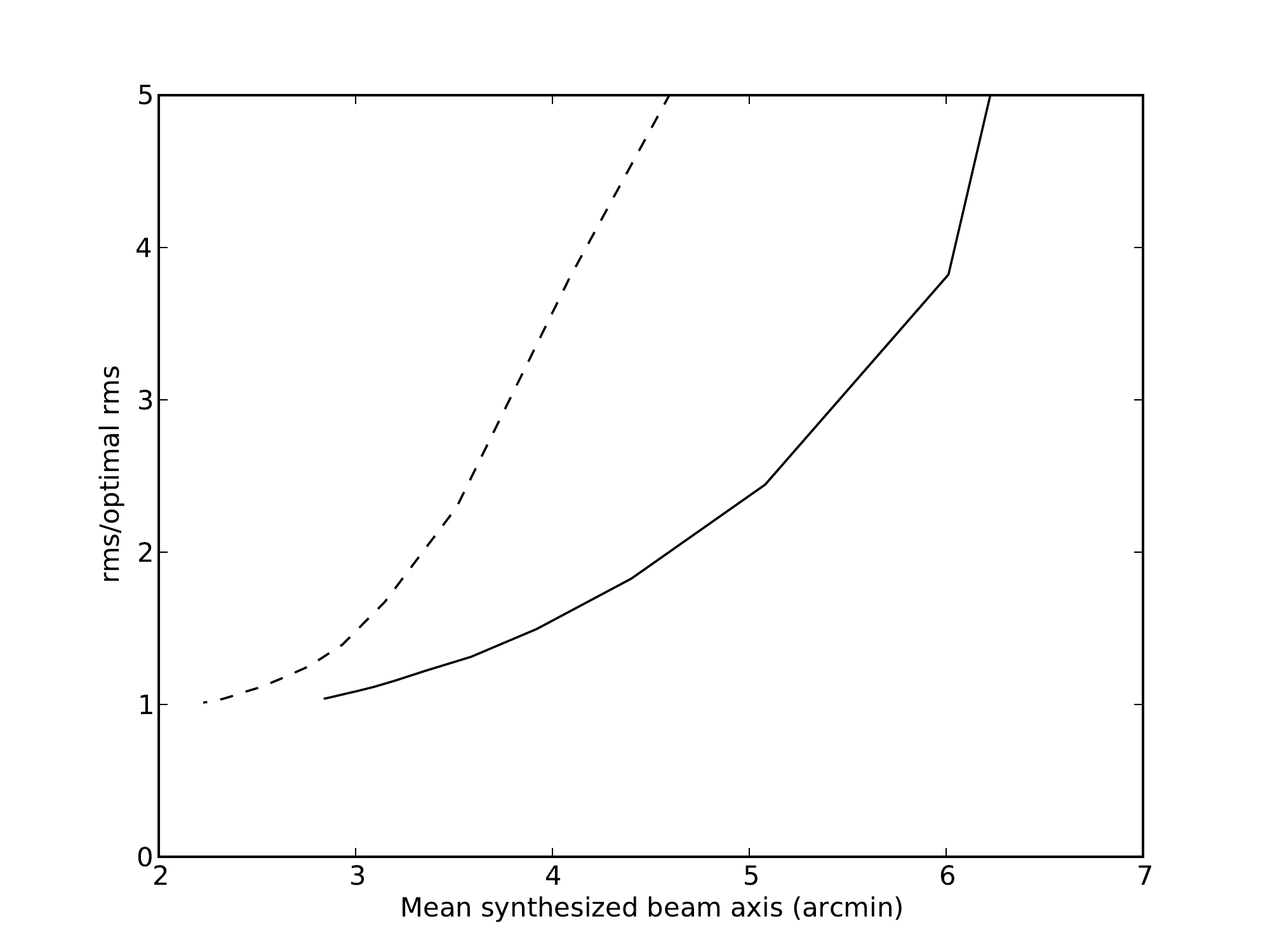}
  \caption{The synthesised beam FWHM versus the flux-density sensitivity normalised by the optimal flux-density sensitivity, where the flux-density sensitivity is the rms of pixels on a map in a region away from the primary beam. Naturally weighting the visibilities provides the optimal flux-density sensitivity but to increase the surface brightness sensitivity a Gaussian $uv$-taper is applied during the imaging procedure in \textsc{aips}. As a result of applying a $uv$-taper the FWHM of the synthesised beam is increased  but the flux-density sensitivity is decreased. The solid line is for the SA and the dashed line is for the SZA. The SA is sensitive to larger angular scales than the SZA. }
  \label{fig:sa-sza-resolution-sensitivity}
\end{figure}

Our Bayesian analysis is performed on the $uv$-data rather than the image and thus avoids problems associated with imaging. We have performed several different analyses of the data and have derived probabilities of detection as well as physical and phenomenological cluster parameters. Each of these analyses takes into account radio sources, receiver noise and the statistics of the primary CMB anisotropies. 

The joint AMI \& SZA analyses was performed with two parametric physical cluster models: the $\beta$-model; and the model described in \citealt{Olamaie_2012}. The cluster mass derived from these different models is comparable, we find $M_{T,200}= 4.1\pm 1.1 \times10^{14}M_{\odot}$ and $M_{T,200}= 5.1\pm 1.2 \times10^{14}M_{\odot}$ for the $\beta$-model and the \cite{Olamaie_2012} model respectively. For this cluster our Bayesian evidences prefer a $\beta$-model ($\Delta Z_{1}(S) =1.1$) and this may be because the shape parameters are allowed to vary in our analysis with the $\beta$-profile which allows it to better fit the data. Whereas we have used the \cite{Olamaie_2012} model with the shape of the pressure profile fixed according to \cite{Arnaud_2010}. As the Bayesian evidences favour a $\beta$-model we quote the values obtained from this model throughout.

We used our Bayesian analysis to blindly search the SZA data for SZ signatures in a box centred on the SZA pointing centre with sides of length 500$\arcsec$ and the most probable candidate is coincident with the cluster discovered in the AMI blind cluster survey. To quantify the significance of detection in the SZA data with a formal Bayesian probability, we have assumed that the cluster number counts of \citealt{Jenkins_2001} describe the expected number of clusters as a function of mass and redshift within the region (as we did for the AMI blind cluster survey). For this blind analysis we obtained a Bayesian probability of detection of $p=0.22$. However, if we make use of our prior knowledge that the cluster does exist at a known position, we derive a Bayesian probability of detection of $p=0.92$. We have performed the same blind analysis on the AMI and AMI \& SZA datasets and obtained Bayesian probability ratios of 4.5$\times 10^{3}$:1 and 5.4$\times 10^{3}$:1 respectively (see Table \ref{McAdam-params}). 

The physical $\beta$-model cluster parameters derived from the SA, SZA and joint AMI \& SZA analyses are given Table \ref{McAdam-params} and Figures \ref{AMICL1E_tri}, \ref{SZACL1E_tri} and \ref{AMI+SZACLsmall}. To aid comparison between the derived parameters Figure \ref{fig:1d-comparison} shows the marginalised parameter probability distributions from these analyses. We derive consistent values for $\beta$ and $r_c$ from all three analyses. The \citealt{Jenkins_2001} prior that is used to jointly constrain $M_{T,200}$ and $z$ (which are degenerate) does give different results for each dataset. From the SZA analysis we obtain a lower $M_{T,200}$ but a higher $z$ than from the SA analysis. The $M_{T,200}$  estimate obtained from the joint analysis is close to midway between the values obtained  from the SZA and SA data. The $z$ probability distribution obtained from the joint analysis closely resembles the probability distribution obtained from the 
SA only analysis but has a slight increase in probability corresponding to the peak of the SZA probability distribution.


The phenomenological analysis of our data gives SZ temperature decrements with magnitudes that correspond to the flux of the SZ decrements on the SZA and SA images. However, unlike simply quantifying the decrement signal in units of receiver noise, our phenomenological analysis accounts for primordial CMB structures and errors in the fluxes of radio sources as well as receiver noise. We constrain the SZ temperature decrement in the SZA data to $\Delta T_0=-140_{-53}^{+60}\,\mu$K which is lower than obtained from the analysis of either the AMI or AMI \& SZA data ($\Delta T_0=-230_{-40}^{+36}\,\mu$K and $\Delta T_0=-170_{-24}^{+24}\,\mu$K respectively), but the errors are large. The derived parameters from the phenomenological analysis  are presented in Table \ref{McAdam-phen-params}. The probability distribution for the key phenomenological parameter, $\Delta T_0$, is shown in Figure \ref{fig:1d-phen-comparison}.

Our Bayesian analysis was also used to search for the second peak in the initial results from the AMI SZ survey (J~03$^{\rm{h}}$~00$^{\rm{m}}$~14.8$^{\rm{s}}$ +26$^\circ$~10$'$~02.6$''$) by blindly search for SZ signatures in a larger $1000\arcsec \times 1000\arcsec$ box centred on the pointing centre. In the SZA image a 2$\sigma_{SZA}$ decrement is visible at this location but in our large area blind analysis of the SZA data this cluster was not detected. The non-detection is unsurprising given the significance of the negative feature. This second decrement lies at 40\% of the SZA primary beam and we would require targeted SZA observations for further analysis.


Both our probability of detection ratio and parameter estimates are obtained without knowledge of the cluster redshift. Further SZ observations would provide additional insights into the large scale structure of this cluster and may help to identify the cause of the slight discrepancy between the magnitude and morphology of the SZ decrement that was detected by both AMI and the SZA. However, perhaps even more useful, would be X-ray or optical observations which would provide complementary data to improve our constraints on the structure of this cluster in addition to a cluster redshift.

\begin{table*}
\caption{The mean derived physical $\beta$-model parameters, Bayesian evidences and the $R$ ratios for our analysis of the AMI, SZA and AMI \& SZA datasets. The blind analyses are performed with a uniform prior on the cluster position whereas the known cluster analysis is performed with a Gaussian prior on the cluster position (see Table \ref{MC_PRIORS}). The known analysis is only performed on the SZA data. For the blind analysis the $\mu_S$ value is calculated from \citealt{Jenkins_2001} and is equal to 0.34 for the AMI analysis and 0.085 for the SZA and the joint AMI \& SZA analyses.}
 \label{McAdam-params}
\begin{tabular}{lcccccccc}
 \hline

Data                           & $M_{T,200}$ & $z$ & $\beta$ & $r_{c}$ & $T_{e,200}$ & Blind  &  $R$ & Known  \\
          & $\times 10^{14}$$M_{\odot}$  & &  & kpc   & keV & ln$\left(\frac{Z_{1}(S)}{Z_{0}}\right)$ & & ln$\left(\frac{Z_{1}(S)}{Z_{0}}\right)$ \\ \hline
AMI & $5.1_{-1.2}^{+1.1}$ & $0.69_{-0.30}^{+0.31}$ & $1.6_{-0.5}^{+0.5}$ & $440_{-220}^{+220}$ & $5.3_{-1.1}^{+1.1}$ & 9.49 & 4.5$\times10^3$ & -- \\ 
\\ 
AMI \& SZA & $4.1_{-1.1}^{+1.1}$ & $0.78_{-0.38}^{+0.40}$ & $1.5_{-0.6}^{+0.6}$ & $500_{-250}^{+260}$ & $4.7_{-1.1}^{+1.1}$ & 11.06 & 5.4$\times10^3$ & --  \\ 
SZA & $2.2_{-0.5}^{+0.4}$ & $0.9_{-0.35}^{+0.36}$ & $1.6_{-0.5}^{+0.5}$ & $460_{-250}^{+250}$ & $3.3_{-0.6}^{+0.6}$ & 1.24 & 0.29 & 2.55 \\ \hline
 \end{tabular}
\end{table*}

\begin{figure*}
  \includegraphics[width=10.5cm,clip=,angle=0.]{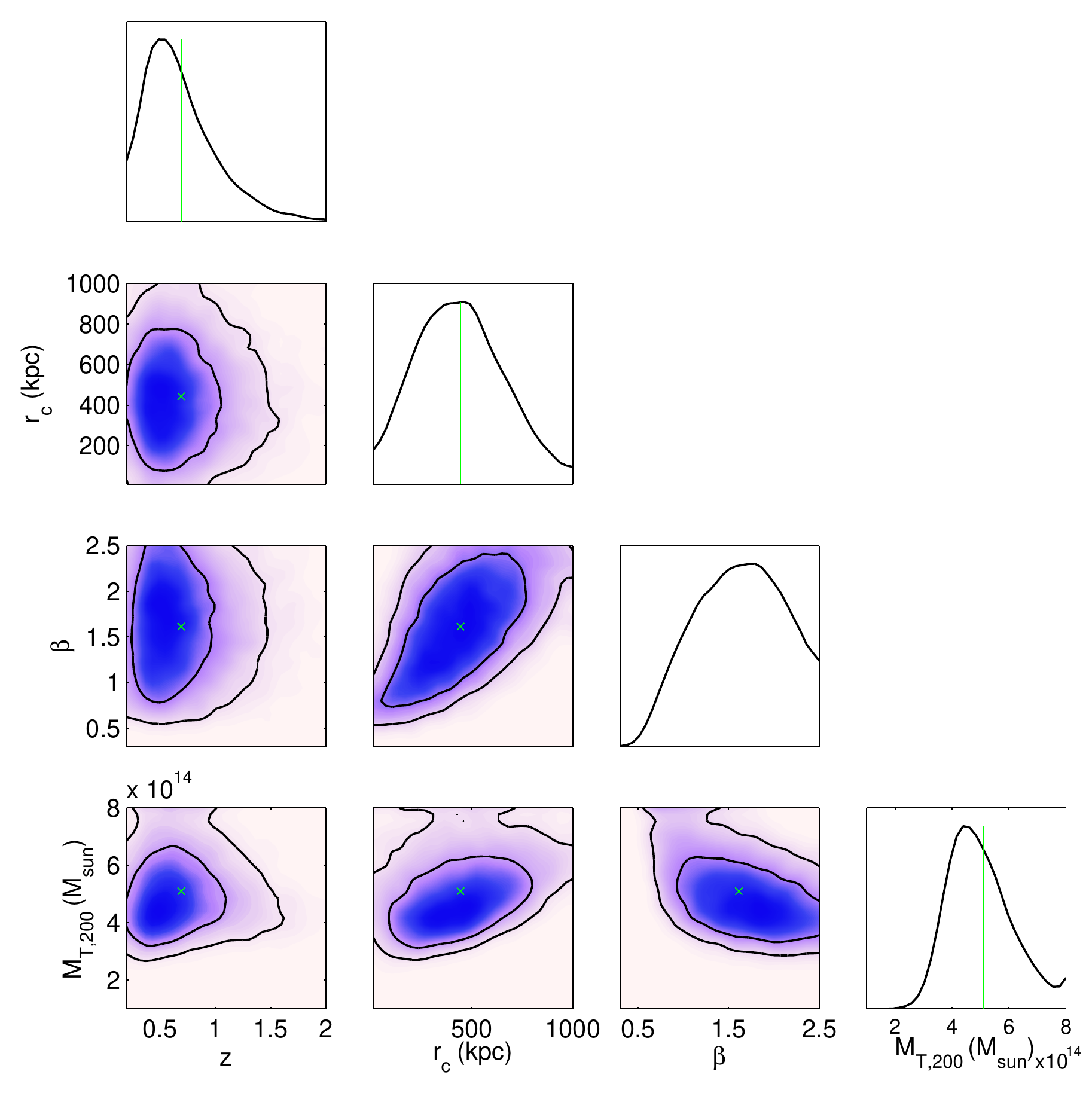}
  \caption{Parameter probability distributions for the physical model analysis of the AMI data. The redshift constraint comes from the \citealt{Jenkins_2001} prior. The crosses show the mean derived parameter values. \label{AMICL1E_tri}}
\end{figure*}

\begin{figure*}
  \includegraphics[width=10.5cm,clip=,angle=0.]{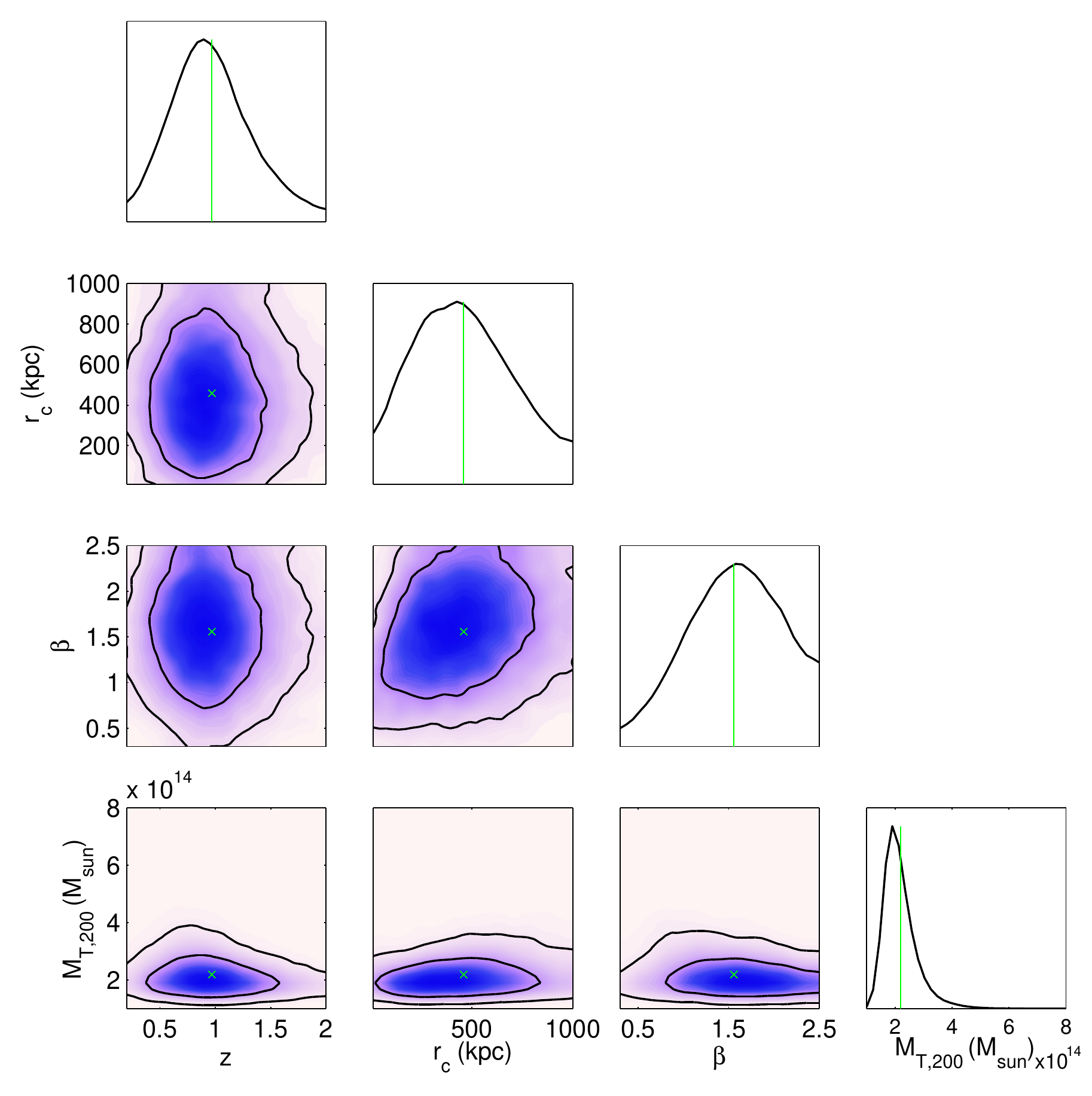}
  \caption{The same as Figure \ref{AMICL1E_tri} but for the analysis of the SZA data.\label{SZACL1E_tri}}
\end{figure*}

\begin{figure*}
  \includegraphics[width=10.5cm,clip=,angle=0.]{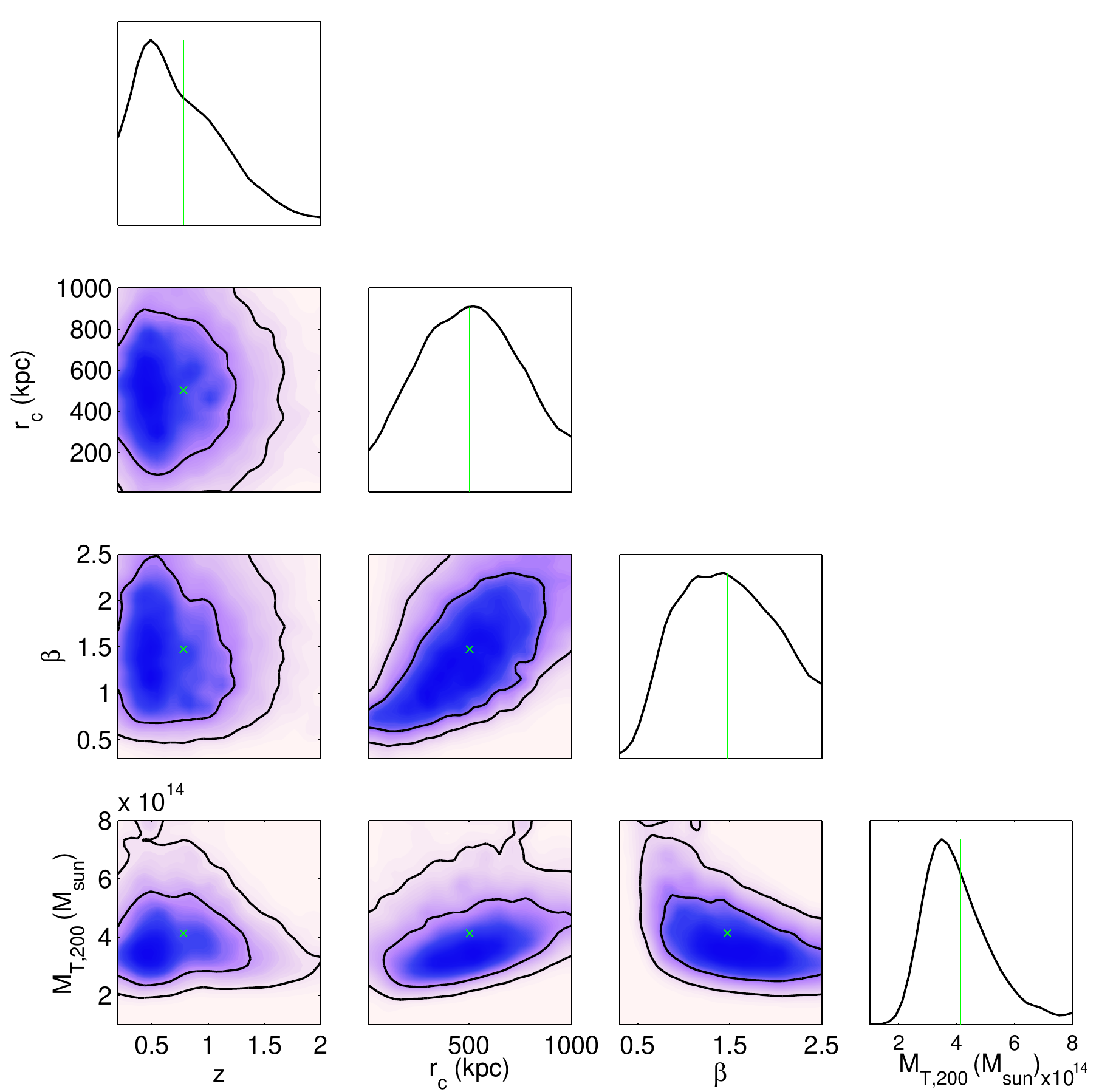}
  \caption{The same as Figure \ref{AMICL1E_tri} but for the joint analysis of the AMI \& SZA data.. \label{AMI+SZACLsmall}}
\end{figure*}

\begin{figure*}
  \includegraphics[width=8.0cm,clip=,angle=0.]{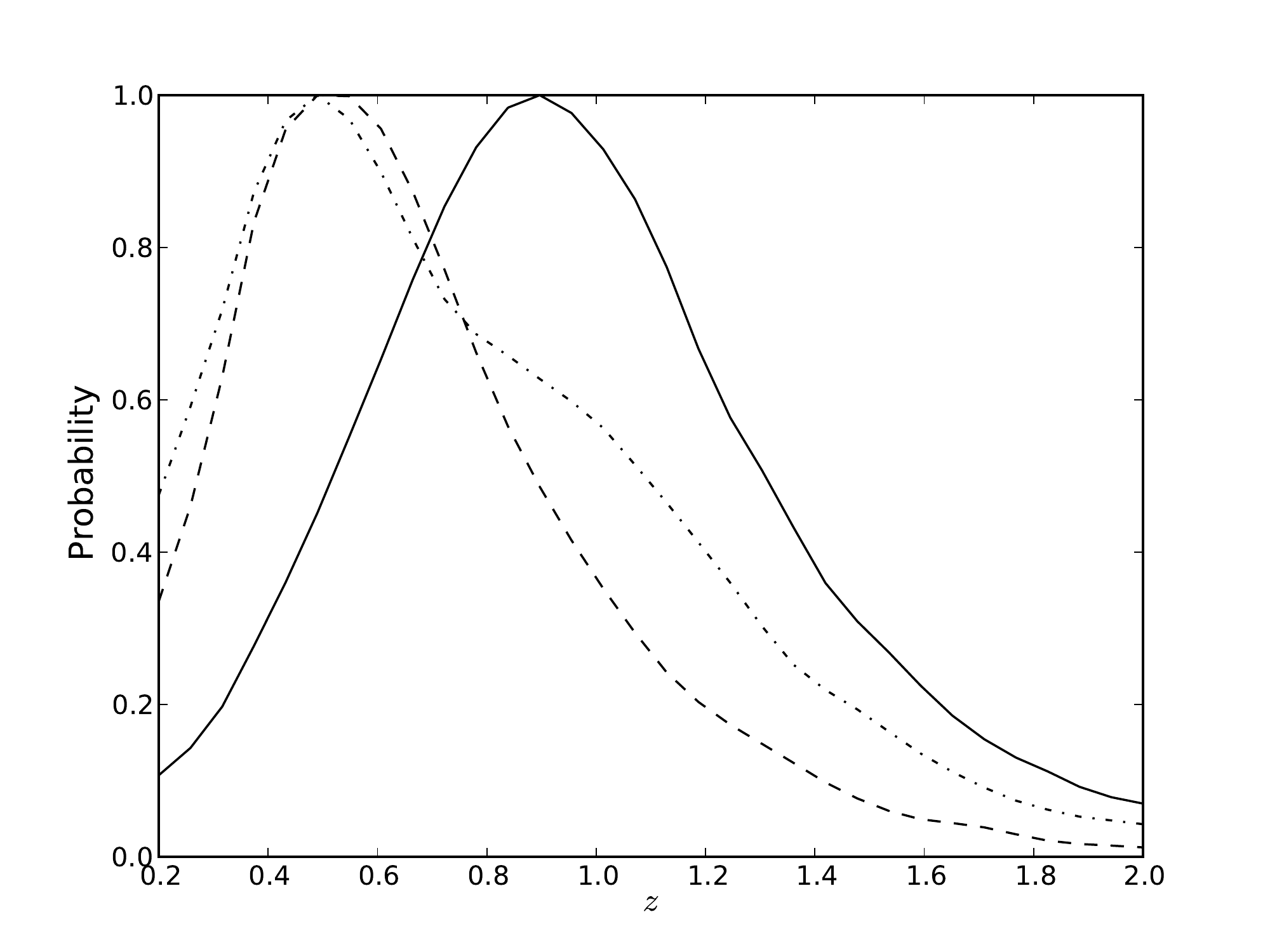} 
  \includegraphics[width=8.0cm,clip=,angle=0.]{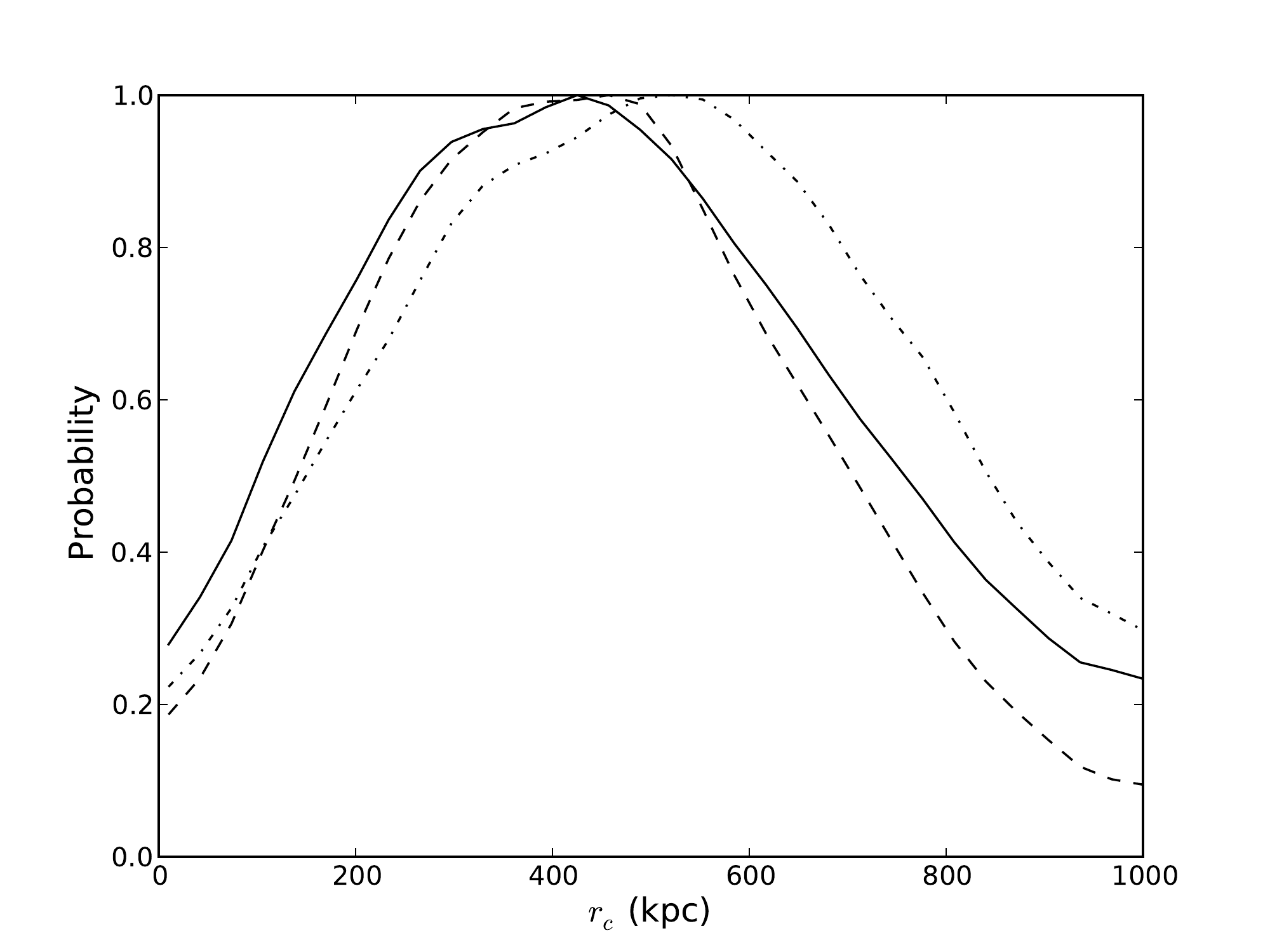}
  \includegraphics[width=8.0cm,clip=,angle=0.]{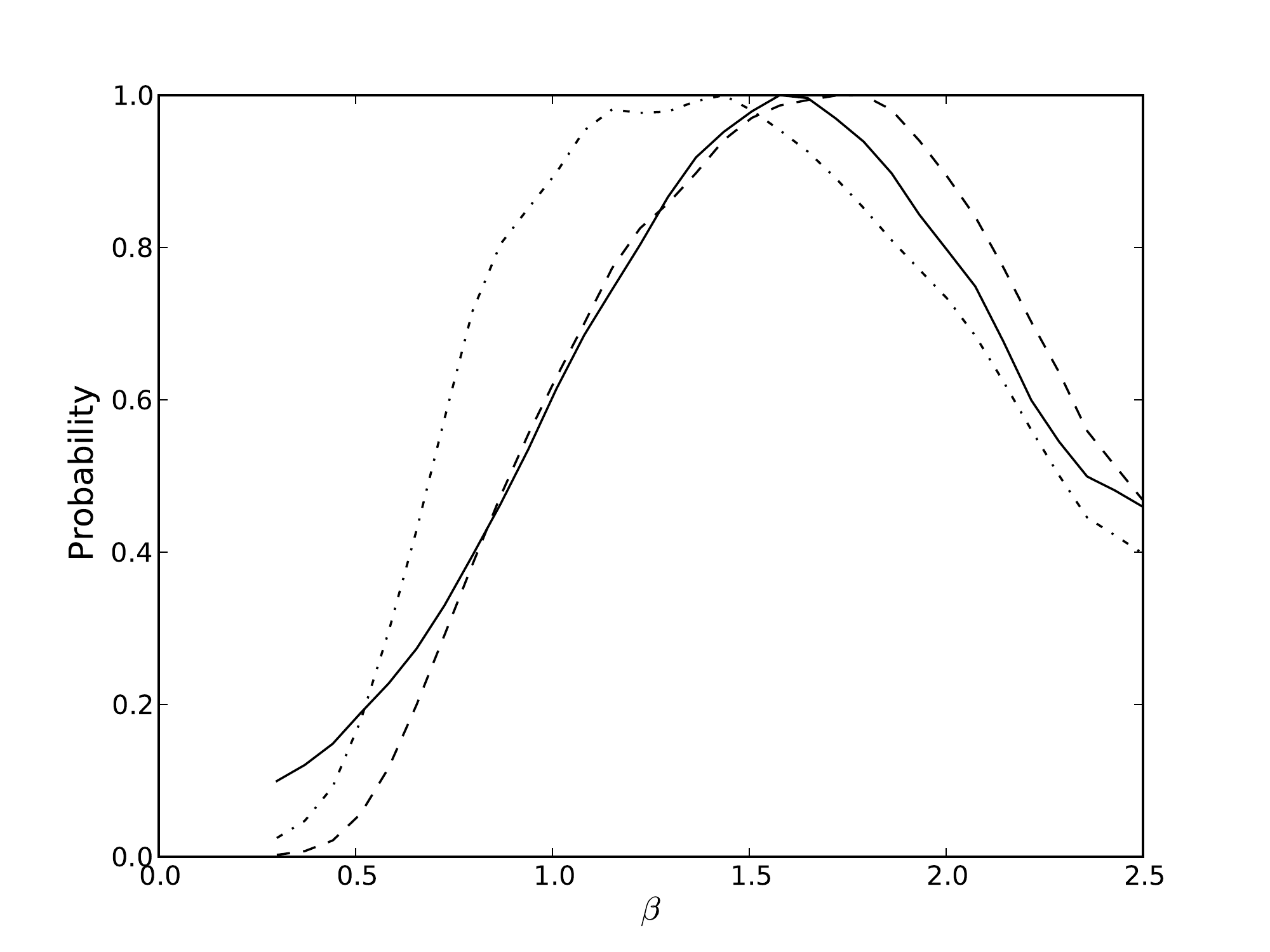}
  \includegraphics[width=8.0cm,clip=,angle=0.]{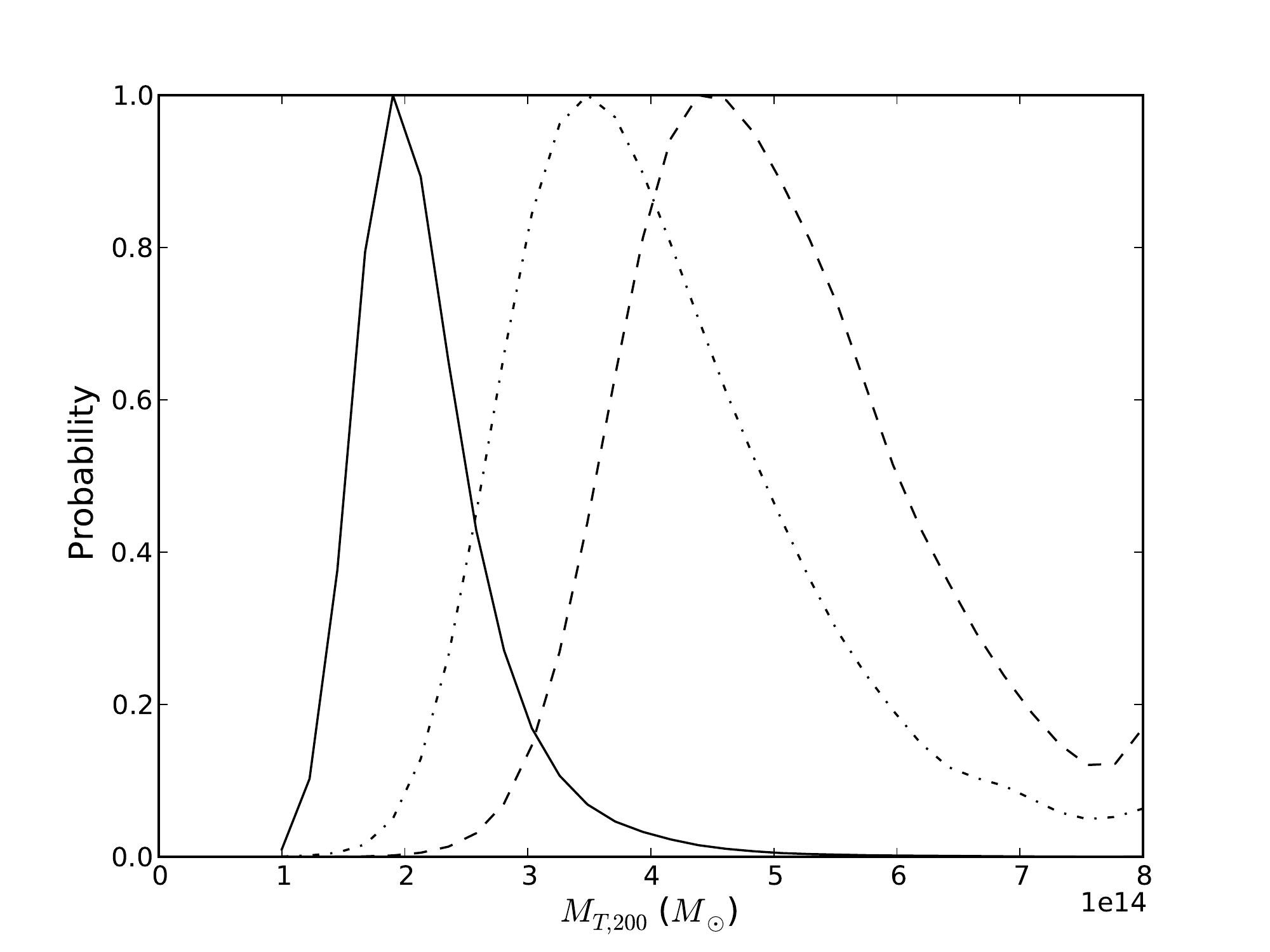}
  \caption{A comparison of the parameter probability distributions in the physical model analysis of the AMI, SZA and AMI \& SZA datasets. The dashed line is the AMI data, the solid line is the SZA data and the dot-dashed line is from the joint analysis of the AMI \& SZA data. \label{fig:1d-comparison}. The redshift constraint comes from the \citealt{Jenkins_2001} prior.}
\end{figure*}

\begin{table}
\caption{The mean derived temperature from the phenomenological analysis of the AMI, SZA and AMI \& SZA datasets.}
 \label{McAdam-phen-params}
\begin{tabular}{lcc}
 \hline
Data          & $\Delta T_{0}$ ($\mu$K) \\ \hline
AMI & $-230_{-40}^{+36}$ \\ 
AMI \& SZA & $-170_{-24}^{+24}$ \\ 
SZA & $-140_{-53}^{+60}$ \\ \hline
 \end{tabular}
\end{table}

\begin{figure}
  \includegraphics[width=8.5cm,clip=,angle=0.]{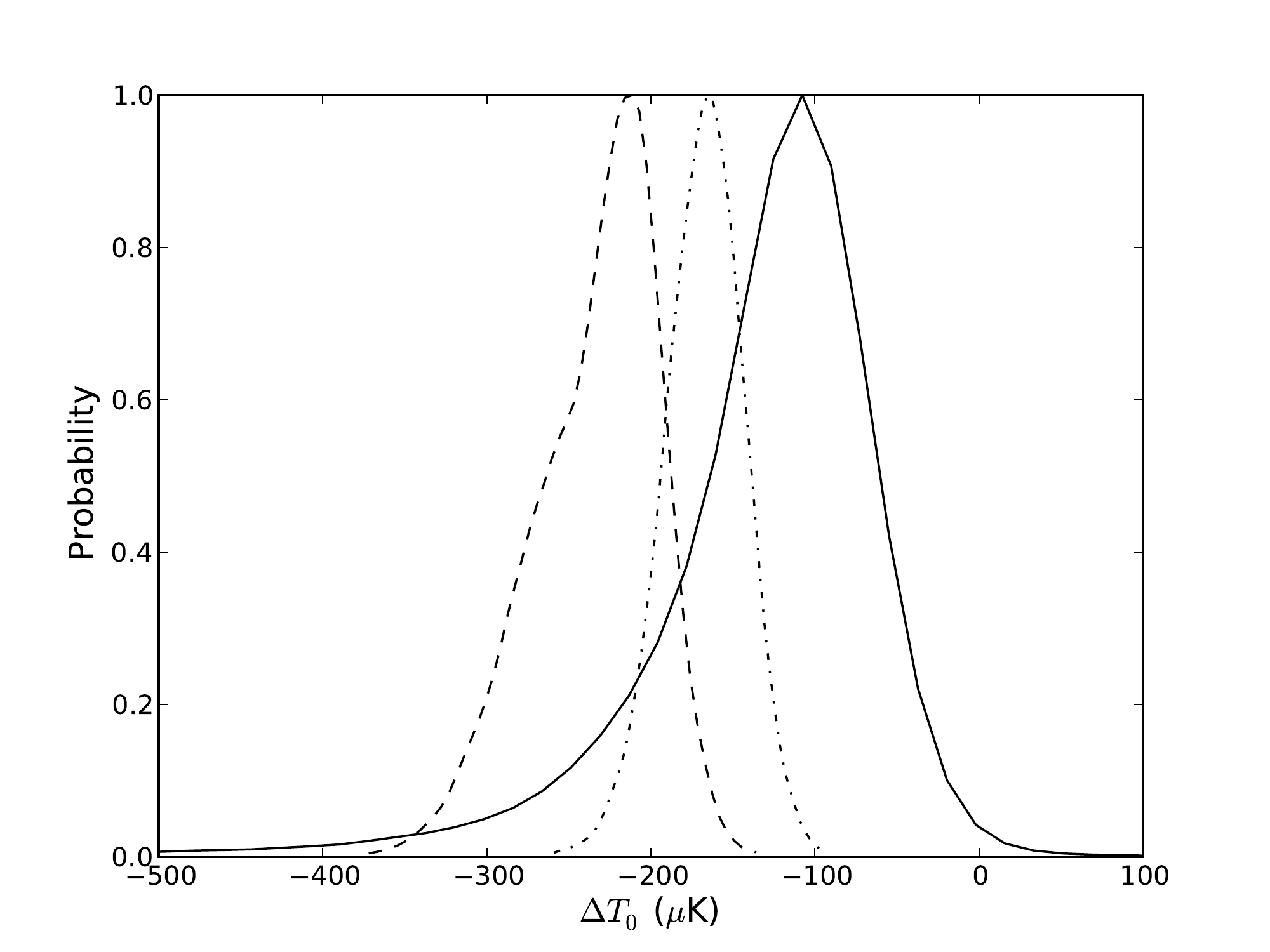}
  \caption{A comparison of the $\Delta T_{0}$ parameter probability distribution in the phenomenological model analysis of the AMI, SZA and AMI \& SZA datasets. The dashed line is the AMI data, the solid line is the SZA data and the dot-dashed line is from the joint analysis of the AMI \& SZA data. \label{fig:1d-phen-comparison}}
\end{figure}

\section{Conclusions}

We have presented SZA follow-up observations of the galaxy cluster system that was discovered in an initial analysis of a twelfth of the AMI blind cluster survey. With analyses of the AMI, SZA and AMI \& SZA datasets we have successfully confirmed the original AMI SZ detection. Our results reveal the need for a careful investigation into the impact of cluster morphology on the SZ structures that are observed with interferometers such as AMI and the SZA and the pipeline that we have developed to jointly analyse data from these two instruments will be utilised for these forthcoming studies. Our conclusions from our analysis of this dataset are the following:

\begin{enumerate}
\item In source-subtracted images made from the SZA follow-up observations the cluster is detected with a peak flux density of 620\,$\mu$Jy (3.9$\sigma_{SZA}$) in the naturally weighted SZA data. By applying a $uv$-taper to the data we increase the magnitude of the detection to 960\,$\mu$Jy (4.0$\sigma_{SZA-TAPER}$).
\item Using the SZA-based Bayesian evidences, we have calculated formal probabilities of detection taking into account point sources, receiver noise and the statistical properties of the primary CMB anisotropy. We find a formal probability of detection ratio of 
0.22:1 when assuming the \citealt{Jenkins_2001} cluster number count, and 12.8:1 when assuming that a cluster exists within our search area. A phenomenological model describing the SZ temperature decrement results in $\Delta T_0=-140_{-53}^{+60}\,\mu$K.
\item Precise knowledge of radio source contamination is important for SZ analysis. We have estimated the 31\,GHz source fluxes using low signal-to-noise estimates from our SZA data. However, deep, high-resolution 31\,GHz observations would improve our knowledge of the source environment and improve the accuracy of our analysis; we look forward to the new 26-36 GHz receivers on the 6.1\,m and 10.4\,m CARMA dishes: improving the ability to handle contaminating sources.
\item We have performed, for the first time, a joint analysis of AMI and SZA data. From our physical cluster model we derive $M_{T,200}= 4.1\pm 1.1 \times10^{14}M_{\odot}$ and from the phenomenological model we obtain  $\Delta T_0= -170\pm24\,\mu$K.
\end{enumerate}

\section{Acknowledgements}

We thank the anonymous referee for providing us with useful comments and suggestions. We thank Cambridge University and STFC for the support of AMI and its operations. We are grateful to the staff of the Cavendish Laboratory and the Mullard Radio Astronomy Observatory for the maintenance and operation of AMI. CR and MPS acknowledge PPARC/STFC studentships. YCP acknowledges the support of a Rutherford Foundation/CCT/Cavendish Laboratory studentship. This work was performed using the Darwin Supercomputer of the University of Cambridge High Performance Computing Service (http://www.hpc.cam.ac.uk/), provided by Dell Inc. using Strate- gic Research Infrastructure Funding from the Higher Education Funding Council for England, and the Altix 3700 Supercomputer at DAMTP, University of Cambridge supported by HEFCE and STFC. We are grateful to Stuart Rankin and Andrey Kaliazin for their computing assistance.

\bsp
\label{lastpage}

\end{document}